\documentclass[10pt,journal,compsoc]{IEEEtran}

\usepackage{amsmath,bm}
\usepackage{graphicx}
\usepackage{multirow}
\usepackage{xspace}
\usepackage[nocompress]{cite}
\usepackage{color}
\usepackage{url}
\usepackage{booktabs}
\usepackage{tablefootnote}

\newcommand{\ie}{\emph{i.e.,}\xspace}
\newcommand{\eg}{\emph{e.g.,}\xspace}
\newcommand{\etal}{\emph{et al.}\xspace}
\newcommand{\etc}{\emph{etc.}\xspace}
\newcommand{\aka}{\emph{a.k.a.}\xspace}

\begin{document}
\title{A Survey of Location Prediction on Twitter\thanks{Accepted for publication at TKDE  https://doi.org/10.1109/TKDE.2018.2807840}}

%
\author{Xin~Zheng, Jialong~Han, and Aixin~Sun%
\IEEEcompsocitemizethanks{
	\IEEEcompsocthanksitem X.~Zheng is with School of Computer Science and Engineering, Nanyang Technological University, Singapore 639798, and SAP Research and Innovation Singapore, SAP Asia Pte Ltd, Singapore 119968.
	 \protect\\
E-mails: xzheng008@e.ntu.edu.sg; xin.zheng@sap.com
\IEEEcompsocthanksitem J.~Han is with Tencent AI Lab, Shenzhen, China.\protect\\
E-mail: jialonghan@gmail.com
\IEEEcompsocthanksitem A.~Sun is with School of Computer Science and Engineering, Nanyang Technological University, Singapore 639798.\protect\\
E-mail: axsun@ntu.edu.sg
\IEEEcompsocthanksitem Corresponding author: Jialong Han.
}
}


\IEEEtitleabstractindextext{%
\begin{abstract}
Locations, \eg countries, states, cities, and point-of-interests, are central to news, emergency events, and people's daily lives.
Automatic identification of locations associated with or mentioned in documents has been explored for decades.
As one of the most popular online social network platforms, Twitter has  attracted a large number of users who send millions of tweets on daily basis.
Due to the world-wide coverage of its users and real-time freshness of tweets, location prediction on Twitter has gained significant attention in recent years.
Research efforts are spent on dealing with new challenges and opportunities brought by the noisy, short, and context-rich nature of tweets.
In this survey, we aim at offering an overall picture of location prediction on Twitter.
Specifically, we concentrate on the prediction of user home locations, tweet locations, and mentioned locations.
We first define the three tasks and review the evaluation metrics.
By summarizing Twitter network, tweet content, and tweet context as potential inputs, we then structurally highlight how the problems depend on these inputs.
Each dependency is illustrated by a comprehensive review of the corresponding strategies adopted in state-of-the-art approaches.
In addition, we also briefly review two related problems, \ie semantic location prediction and point-of-interest recommendation.
Finally, we make a conclusion of the survey and list future research directions.
\end{abstract}

}

\maketitle

\IEEEdisplaynontitleabstractindextext

\IEEEpeerreviewmaketitle

\IEEEraisesectionheading{
	\section{Introduction}
    \label{sec:introduction}
    }

\IEEEPARstart{T}{he} last decade has witnessed an unprecedented proliferation of online social networks.
Those include general-purpose platforms like Twitter and Facebook, location-based ones like Foursqure and Gowalla, photo-sharing sites like Flickr and Pinterest, as well as other domain-specific platforms such as  Yelp and LinkedIn.
On these platforms, users may establish online friendship with others sharing similar interests.
Users may also share with online friends their daily lives in forms of texts, photos, videos, or check-ins.

Among all online social networks, Twitter is characterized by its unique way of following friends and sending posts.
On the one hand, Twitter friendships are not necessarily mutual.
For example, users may ``follow'' celebrities without requiring them to follow back.
On the other hand, textual posts on Twitter, \aka tweets or microblogs, are limited to 140 characters.
Users are encouraged to post frequently but casually about anything, such as moods, activities, opinions, local news, \etc

Users, online friendships, and tweets make Twitter a virtual online world.
This virtual world intersects with the real world, where locations acting as intermediate connections.
Twitter users have long-term residential addresses.
Their home locations cause them to notice, get interested, and tweet news or events around their daily activity regions.
With increasing popularity of GPS-enabled devices such as smartphones and tablets, users may casually attach real-time locations when sending out tweets.
Users may also mention locations in their tweets, \eg cities they previously lived in, or restaurants they want to try.
In this survey, we concentrate on the above three types of Twitter-related locations, namely \textbf{\emph{user home location}}, \textbf{\emph{tweet location}}, and \textbf{\emph{mentioned location}}.
Knowing physical locations involved in Twitter helps us to understand what is happening in real life, to bridge the online and offline worlds, and to develop applications to support real-life demands, among many applications.
For example, we can monitor public health of residents~\cite{cheng2010you}, recommend local events~\cite{yuan2013and} or attractive places~\cite{noulas2012mining} to tourists, summarize regional topics~\cite{rakesh2013location}, and identify locations of emergency~\cite{ao2014estimating} or even disasters~\cite{lingad2013location}.

Although Twitter users may casually reveal locations either manually or with the help of GPS, location information on Twitter are far from complete and accurate.
Cheng \etal~\cite{cheng2013content} find that only 21\% of users in a U.S. Twitter dataset provide residential cities in their profiles, while 5\%  give coordinates of their home addresses.
Despite the low availability, Hecht \etal~\cite{hecht2011tweets} report that self-declared home information in many user profiles are inaccurate or even invalid.
Hecht \etal~\cite{hecht2011tweets} and Ryoo \etal~\cite{ryoo2014inferring} observe that only 0.77\% and 0.4\% of tweets have location information attached in their datasets, respectively.
Similar percentages are also reported by Bartosz \etal~\cite{hawelka2014geo} and Priedhorsky \etal~\cite{priedhorsky2014inferring}.
Therefore, completing Twitter-related locations acts as the prerequisite for many other studies and applications, and is worth careful investigation.

The problem of predicting locations associated with objects has been termed as geolocation or geocoding, and studied for Wikipedia~\cite{Wing:2011:SSD:2002472.2002593,wing2014hierarchical,roller2012supervised}, web pages~\cite{amitay2004web,Zong2005JCDL}, and general documents~\cite{woodruff1994gipsy}.
The recognition and disambiguation of mentioned entities\footnote{A named entity is a real-world object; examples are persons, organizations, or locations.} in formal documents, or entity recognition~\cite{nadeau2007survey} and linking~\cite{DBLP:journals/tkde/ShenWH15}, are also extensively investigated for decades.
Various text processing techniques have been proposed to address these problems.
Intuitively,  recognition and disambiguation of Twitter-related locations should also depend heavily on tweet texts.
Users living in certain cities may discuss local landmarks, buildings and events, possibly with dialects or slang.
Tweets sent out from certain locations may explicitly mention them in the text, or implicitly include some relevant words.
However, the characteristics of Twitter pose emerging challenges for these existing research problems in new problem settings.
On the one hand, users often write tweets in a very casual manner.
Acronyms, misspellings, and special tokens make tweets noisy, and techniques developed for formal documents are error-prone on tweets.
The limit of 140-character also makes tweets short, which may not be easily understood by readers who are unaware of tweets' context.
On the other hand, compared with formal documents, Twitter users contribute their online friendships and profiles explicitly.
They also intentionally or unintentionally attach geo-tags to tweets.
The richness of contextual information on Twitter enables new opportunities to relieve  aforementioned challenges.

Given the above significance, necessity, challenges, and opportunities, Twitter-related location prediction problems have received much attention in the literature, and even been proposed as one of the shared tasks in the 2nd Workshop on Noisy User-generated Text (W-NUT)~\footnote{The workshop also provides an evaluation dataset which we call W-NUT(\url{http://noisy-text.github.io/2016/index.html}).}.
To the best of our knowledge, no previous survey focuses extensively on exactly the same scope.
Imran \etal~\cite{imran2014processing} have done a comprehensive study on tracking and analyzing mass emergency with social media data.
Their focus is multifaceted, which not only involves locations but also has temporal and event aspects.
Melo \etal~\cite{TGIS:TGIS12212} review various techniques for geolocating ordinary documents, but the unique challenges and opportunities of Twitter are not touched.
Ajao \etal~\cite{DBLP:journals/jis/AjaoHL15} conduct a smaller scale survey which addresses the most similar scope as we are aware of.
However, they only clarify possible input and output of location prediction problems on Twitter.
Detailed techniques are discussed with minimal efforts.
Nadeau \etal~\cite{nadeau2007survey} and Shen \etal~\cite{DBLP:journals/tkde/ShenWH15} concentrate on named entity recognition and linking, respectively.
They are  related to one of the three problems in this survey, \ie mentioned location prediction.
Besides, their focuses are on general entities and documents, while we specially target the intersection of the location domain and Twitter platform.

In this survey, we aim at completing an overall picture of location prediction problems on Twitter.
In Section~\ref{sec:problemOverview}, we brief the input, output, and evaluation metrics of Twitter-based location prediction.
In Sections~\ref{sec:homeLocation},~\ref{sec:tweetLocation} and~\ref{sec:locationLinking}, we detail previous efforts on each problem.
By highlighting the role of each input, we systematically summarize essentials of previous works on each prediction problem.
In Section~\ref{sec:other}, we  brief two additional location-related problems.
Though attracting less attention or not as relevant, these two problems complement the three major problems and the scope of this survey.
Finally, we conclude the survey and discuss future research directions.

\section{Problem Overview}
\label{sec:problemOverview}

\begin{figure}
  \centering
  \includegraphics[width=0.95\columnwidth]{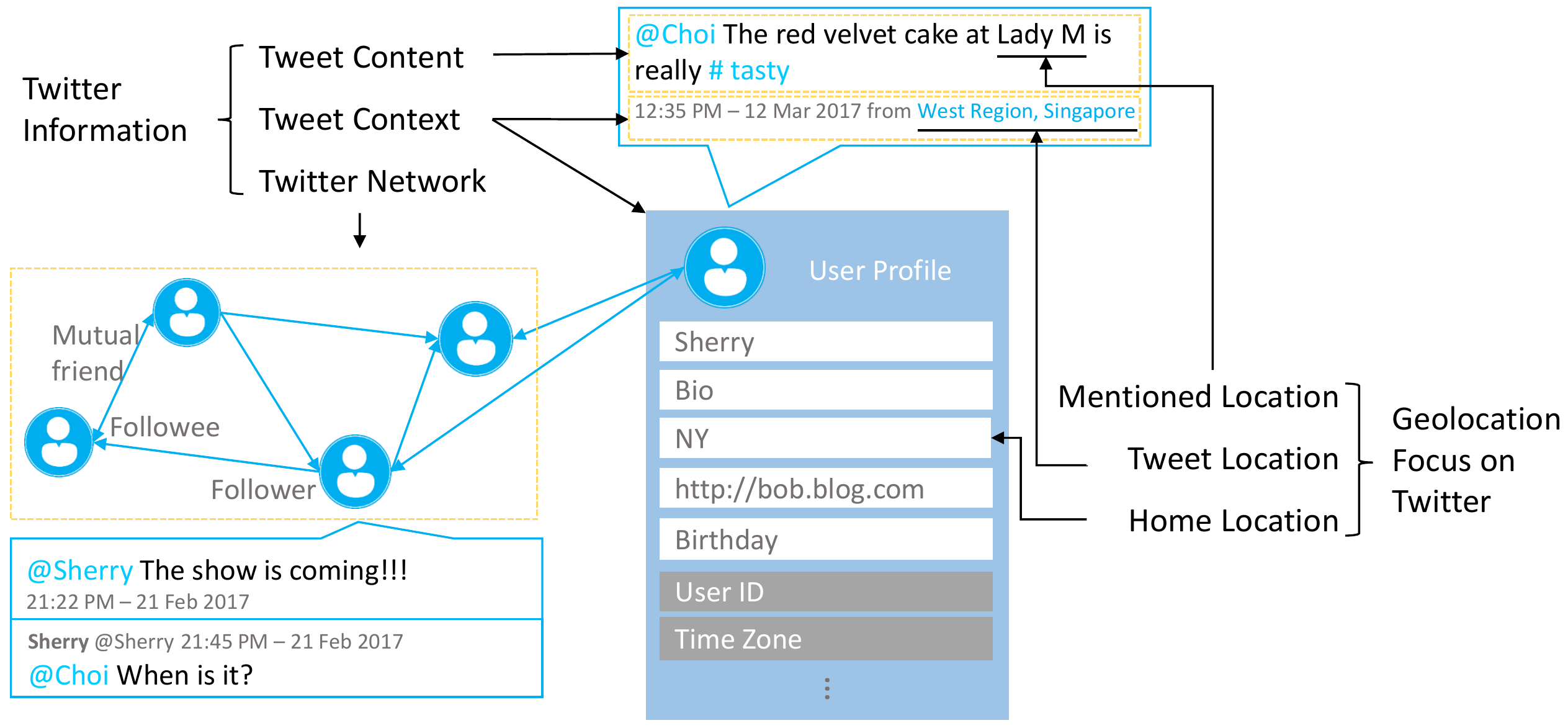}
  \caption{An illustration of tweet content, tweet context, and Twitter network, and the three types of locations: home location, tweet location, and mentioned location in Twitter.}
  \label{fig:example}
\end{figure}

This survey focuses on location prediction problems on Twitter.
In this section, first, we give an overview of the Twitter platform.
By introducing Twitter usage from an ordinary user's point of view, we summarize Twitter dataset from three perspectives \ie content, network, and context.
Next, we discuss three geolocation problems of general interest.
Those prediction problems rely on the above information as major input.
Finally, we briefly review evaluation metrics for the aforementioned prediction problems.

\subsection{An Overview of Twitter}
\label{ssec:input}

As one of the most popular online social network, Twitter constantly accumulates large volume of heterogeneous data at a high velocity.
Those include 1) short and noisy tweets posted by users, 2) a massive Twitter network established among users, and 3) rich types of contextual information for both users and tweets.
Such information serves as input and enables the study of a few geolocation problems.
In this section, we briefly describe the three types of information.

\subsubsection{Tweet Content}
\label{sssec:content}

A \emph{tweet} is a piece of user-generated text with its length up to 140 characters.
It may describe anything a user wants to post, \eg her mood or events happening around her.
Besides original posts, a user may also \emph{retweet} others' tweets she reads.
Tweets and retweets from a user will be pushed to her \emph{followers'} (see definition in Section~\ref{sssec:friendship}) Twitter interface for them to read.
When composing tweet contents, a user may include \emph{hashtags}, which are words or unspaced phrases starting with ``\#''.
Finally, one can also \emph{mention} another user's name by a preceding ``@'' in tweet content.
A mentioned user will be notified, and may start a \emph{conversation} with the mentioning user through subsequent mentions.

\subsubsection{Twitter Network}
\label{sssec:friendship}

Besides posting tweets, a user may subscribe others' tweets by \emph{following} them.
If user $u_i$ follows $u_j$, we call $u_i$ the \emph{follower}, and $u_j$ the \emph{followee}.
Note that following relationships are unidirectional, \ie $u_i$ following $u_j$ does not necessarily mean $u_j$ following $u_i$.
When the direction of a following relationship is not the major concern, we regard $u_i$ and $u_j$ as \emph{friends}.
If it happens that $u_i$ and $u_j$ follow each other, we say $u_i$ and $u_j$ are \emph{mutual friends}.
We refer to all `following' relationships as \emph{Twitter friendship}, or \emph{friendship} when the context is clear.

Note that Twitter friendship does not imply  friendship in real life.
It is often the fact that celebrities do not follow back most of their ordinary followers.
Moreover, even two distant strangers may become mutual friends by chance.
However, it is observed that friends in real life tend to mention each other frequently online~\cite{mcgee2011geographic,mcgee2013location,compton2014geotagging,jurgens2013s}.
When introducing the studies on clues that imply real-life friendship, we  consider both following and mentioning actions between Twitter users in a uniform manner, and refer to the resulted network as \emph{Twitter network}.

\subsubsection{Tweet Context}
\label{sssec:context}

A tweet is more than a piece of short text.
When a tweet is sent out, it is attached with its posting \emph{timestamp}.
Moreover, with the prevalence of GPS-enabled devices like smartphones and tablets, users may optionally publish their current locations as \emph{geo-tags}\footnote{Geo-tags may be in the form of point-of-interests (\eg a hotel or a shopping mall) or simply geographical coordinates (latitudes and longitudes).} on tweets.
Finally, users may complete their profiles to include information like home cities, timezones, and personal websites.
We note that all above information provide context helping us better understand tweets.
A user's daily-life tweets can be interpreted more precisely, if all such information are available.
Because timestamps, geo-tags, and user profiles serve as contextual information for tweets, we refer to them as \emph{tweet context}.

\subsection{Location Prediction Problems on Twitter}
\label{ssec:output}

In this survey, we focus on predicting three types of Twitter-related locations, namely \emph{home location}, \emph{tweet location}, and \emph{mentioned location}.
For each type of location, we give its definition and show how it is represented.
We also briefly discuss how to set up ground truth for each task.

\subsubsection{Home Location Prediction}
\label{sssec:homeLocation}

Home locations refer to Twitter users' long-term residential addresses.
The prediction of home locations enables various applications, \eg local content recommendation, location-based advertisement, public health monitoring, and public opinion polling estimation.
According to specific requirements of applications, home locations may be represented at different levels of granularity.
Generally, there are three categories of home location granularity:
\begin{itemize}
  \item \textbf{Administrative regions}, \ie  countries, states, or cities where  users stay.
  \item \textbf{Geographical grids}, \ie the earth is partitioned into cells of equal or varying sizes\footnote{Equal-sized cells are achieved by uniformly binning latitudes and longitudes~\cite{Wing:2011:SSD:2002472.2002593}. The major drawback is that rural areas are over-represented at the expense of urban areas. Therefore, quad-tree~\cite{yamaguchi2014online} or $k$-dimensional tree ($k$-d tree)~\cite{roller2012supervised,DBLP:conf/coling/HanCB12,wing2014hierarchical} are adopted to achieve varying-sized cells with better resolutions on populated areas.}, and home locations are represented by the cells they fall in.
  \item \textbf{Geographical coordinates}, \ie homes are represented by their latitudes and longitudes. Coordinates may be self-reported or converted from administrative regions or cells by taking their centers.
\end{itemize}
Ground truth home locations may be collected from users' self-declared profiles.
For example, in Figure~\ref{fig:example}, the user reports that she lives in NY (New York).
Due to possible privacy concerns, empty and noisy information appears in user profiles. Some studies also aggregate geo-tags attached with users' tweets as their ground truth home locations.
Possible aggregating approaches include:
\begin{itemize}
  \item The most frequent city involved in the geo-tags.
  \item The first valid geotag, and convert it to an administrative region, a grid, or coordinates.
  \item The geometric median\footnote{The geometric median of a point set $S$ is the point in $S$ which has minimal average distance to the other points.} of the geo-tags.
\end{itemize}
For the sake of evaluation, a uniform level of granularity should be decided and fixed for an application.
However, to achieve maximum coverage of ground truth, user profiles and geo-tag aggregations could be utilized in combination.

\subsubsection{Tweet Location Prediction}
\label{sssec:tweetLocation}

Tweet location means the place where a tweet is posted.
By inferring tweet locations, we may draw a more complete picture of a user's mobility.
Different from home locations, which are collected from both user profiles and geo-tags, tweet locations are generally based on geo-tags of tweets.
Due to the original views of tweet locations, point-of-interests (POIs in short) or coordinates are broadly adopted as representations of tweet locations, instead of administrative regions or grids.

\subsubsection{Mentioned Location Prediction}
\label{sssec:locationMention}

When writing tweets, users may mention the names of some locations in tweet contents.
Mentioned location prediction may facilitate better understanding of tweet contents, and benefit applications like location recommendation and disaster \& disease management.
In this survey, we involve two sub-tasks of mentioned location prediction:
\begin{itemize}
  \item \textbf{Mentioned location recognition}, \ie extract text fragments in a tweet that refer to location names.
  \item \textbf{Mentioned location disambiguation}, \ie identify what locations those fragments refer to by resolving them to entries in a location database.
\end{itemize}
Due to the inherent noise and ambiguity of tweet language, ground truth of mentioned locations largely rely on human annotations.
To represent location mentions in tweets, BIO or BILOU\footnote{BIO stands for the \textbf{B}eginning, \textbf{I}nside, and \textbf{O}utside of a location mention in a sentence. BILOU additionally annotates the \textbf{L}ast word of a multi-word mention, as well as all \textbf{U}nit-length mentions.} labeling schemes are widely adopted.
For both sub-tasks, the granularity of locations involve both administrative regions and POIs.
When a pre-defined location database is employed, the granularity generally respects that of the database.

\subsection{Twitter Inputs for Location Prediction Problems}
\label{ssec:inputs2problems}

All the three types of information on Twitter, \ie content, network, and context, are commonly adopted to solve the three location prediction problems, \ie the prediction of home location, tweet location and mentioned location. This is because multiple data source could help to enrich the available information, so that to relieve data sparsity issue on Twitter. However, for different geolocation problems, the ways to utilize the input data are different. We will discuss the differences at the end of each section.

\subsection{Evaluation Metrics}
\label{ssec:metrics}

In this section, we review common evaluation metrics adopted in the literature.
Depending on the representations of predicted and ground-truth locations that are fed to the evaluation stage, common metrics could be categorized as \emph{distance-based} or \emph{token-based}.
In the distance-based point of view, locations are represented by their geographical coordinates.
Token-based metrics treat locations as discrete symbols, \eg country, city, grid, POI.
Next, we formulate both of them and demonstrate their usage scenarios.

\subsubsection{Distance-based Metrics}
\label{sssec:coordinateMetrics}

In home location or tweet location prediction, we aim at making predictions for each user or tweet.
For unified notations, let $s$ be a user or tweet, and $S$ be the set of all users or tweets for prediction.
A system is expected to predict a location $l(s)$ for each $s$.
The prediction $l(s)$ is expected to coincide with or be close to the ground truth location $l^{*}(s)$.
Whatever granularity we adopt, all ground-truth and predicted locations could be converted to coordinates.
\textbf{Error Distance} (\textbf{ED} for short) is then defined as the Euclidean distance between ground-truth and predicted coordinates:
\[
\mathrm{ED}(s)=\text{dist}(l(s), l^{*}(s)).
\]
Since evaluations are conducted on a collection of users or tweets, we may take the mean or median of all error distances to end up with corpus-level metrics.
This results in \ie \textbf{Mean Error Distance} and \textbf{Median Error Distance}:
\begin{align*}
\mathrm{MeanED} &= \frac{1}{|S|} \underset{s\in S}{\sum} \text{dist}(l(s), l^{*}(s)),\\
\mathrm{MedianED} &= \underset{s\in S}{\mathrm{median}} \{\text{dist}(l(s), l^{*}(s))\}.
\end{align*}
When wildly inaccurate predictions occur, Median Error Distance is usually less sensitive than Mean Error Distance.
Therefore, Mean Error Distance is preferred by some studies.
Instead of Mean Error Distance, some studies~\cite{schulz2013multi}, though very few, employ \textbf{Mean Squared Error} as below:
\[
\mathrm{MSE} = \frac{1}{|S|}\underset{s\in S}{\sum} \text{dist}^2 (l(s), l^{*}(s)).
\]
The only difference between Mean Squared Error and Mean Error Distance is the former takes square of Error Distance.

Besides Mean and Median Error Distance, there is another widely-adopted corpus-level metric called \textbf{Distance-based Accuracy}, or \textbf{Acc@d} for short.
Given a predefined threshold $d$ of error distance, any prediction whose error distance does not exceed $d$ is regarded as ``tolerably correct''.
The Acc@d metric over the corpus is then defined as the proportion of tolerably correct predictions:
\begin{equation}
\mathrm{Acc@d} = \frac{ | \{s \in S: \mathrm{ED}(s)\leq d \} |} {|S|}.
\end{equation}
The commonly adopted distance threshold $d$ is 100 miles, or 161 km~\cite{li2012towards,han2013stacking}.

\subsubsection{Token-based Metrics}
\label{sssec:tokenMetrics}

Alternatively, token-based metrics treat locations as discrete symbols, \eg country, city, grid, POI.
Though geographical information is not taken into consideration, token-based metrics allow for more general usage scenarios.

For the three geolocation problems, the simplest token-based metric is Accuracy.
Let $l(s)$ and $l^{*}(s)$ be the predicted and ground-truth locations for a user, a tweet, or a recognized location mention $s$.
Note that their administrative-region or POI representations are kept.
A prediction is deemed correct only if it coincides with the ground-truth.
\textbf{Accuracy} is then defined as the ratio of correct predictions within $S$:
\[
\mathrm{Acc} = \frac{ | \{s \in S: l(s)=l^{*}(s) \} |} {|S|}.
\]

In some cases, a system may give a ranking list $L(s)$ of predicted locations instead of one.
A straightforward approach is to treat the top location as the only prediction and resort to Accuracy.
However, this approach ignores other predictions in the list, which may also be useful when fed to downstream applications or users.
In light of this, \textbf{Ranking-based Accuracy}, or \textbf{Acc@k} is designed.
A ranking list is considered ``correct'' if the ground-truth location lies within the top-k results $L_k(s)$.
Acc@k is then defined as the proportion of ``correct'' lists:
\[
\mathrm{Acc@k} = \frac{| \{s \in S: l^{*}(s)\in L_k(s) \} |} {|S|}.
\]

Finally, we note that the geolocation systems may not be able to make predictions in some cases.
For example, in home and tweet location predictions, some systems cannot assign locations if insufficient information is given~\cite{compton2014geotagging,flatow2015accuracy,schulz2013multi}.
In mentioned location disambiguation, systems may not find appropriate entry for a given location mention.
In such cases, Precision, Recall and $F_{1}$ are adopted as metrics.
Given a user, a tweet, or a recognized location mention $s$, let $l(s)=\text{null}$ if the system cannot make any prediction.
The \textbf{Precision} over the evaluation corpus $S$ is defined as the ratio of correct predictions among all predictions:
\[
\mathrm{Precision} = \frac{ | \{s \in S: l(s)=l^{*}(s) \} |} { | \{s \in S: l(s)\neq \text{null} \} |}.
\]
Meanwhile, \textbf{Recall} is defined similarly as Accuracy, \ie
\[
\mathrm{Recall} = \frac{ | \{s \in S: l(s)=l^{*}(s) \} |} {|S|}.
\]
After Precision and Recall are defined, \textbf{$F_{1}$} is the harmonic mean of Precision and Recall:
\[
{F_{1}} = \frac{2\times Precision\times Recall}{Precision + Recall}.
\]

Finally, we note that Precision, Recall and $F_{1}$ are applicable and are actually widely adopted for mentioned location recognition.
When evaluating location recognition results, mentioned fragments should be regarded as ``tokens''.
A predicted fragment is deemed correct if its left and right boundaries coincide with those of a ground-truth fragment, respectively.
Precision is then defined as the ratio of correctly predicted fragments over all predicted fragments.
Recall is the proportion of correctly predicted fragments among all ground-truth fragments.
Accordingly, their harmonic mean is defined as $F_{1}$.

\section{Home Location Prediction}
\label{sec:homeLocation}

Knowing home locations of Twitter users enables many applications, such as local content recommendation, location-based advertisement, public health monitoring, public opinion polling, \etc
However, because it is optional for Twitter users to complete their profiles, Twitter users' home locations are mostly absent or noisy.
Therefore, many research efforts have been spent on predicting users' home locations.
In most studies, home locations are predicted at city-level, and sometimes at state or country level.
In this section, we detail them based on different inputs, namely tweet content, Twitter network, and tweet context.
Note that many studies simultaneously involve multiple inputs, especially the first two.
In this case, they will be mentioned multiple times, where assumptions and techniques regarding different inputs are discussed in the corresponding subsections.

\subsection{Inference based on Tweet Content}
\label{ssec:tweetContent3}

Users' home locations could be casually revealed by certain words in tweet content.
For example, people in Houston would talk about Houston Rockets more than users in New York.
Residents from Texas usually use dialect ``howdy'' and those from Philadelphia often call themselves ``phillies''.
Thus, the underlying challenge for content-based home location prediction is to precisely link users to locations via those indicative words.

Previous studies on content-based home location prediction could be divided in two classes: word-centric and location-centric. Word-centric method is to estimate the probability of a location $l$ given words $w$ in text, or $p(l|w)$; while location-centric method focuses on the probability of generating a tweet $d$ at a given location $p(d|l)$.
Next, we will detail the two kinds of studies respectively.

\subsubsection{Word-Centric Methods}
\label{sssec:wordcentric}

In the beginning of Section~\ref{ssec:tweetContent3}, we mentioned two examples about location-indicative words in users' tweets.
Word-centric methods aim at identifying and exploiting such words to predict users' home locations.
Not all words are location-indicated.
For example, words like ``downtown'' and ``OMG'' are used everywhere on Twitter.
Therefore, only \emph{local words}, \ie words that show strong locality, should be involved.
Besides, the location information implied by local words, or their \emph{spatial word usage}, should be learnt from data before making predictions.
Next, we describe how both tasks are achieved in the literature.

\subsubsection*{Identifying Local Words}
\label{sssec:localWords}

In information retrieval literature, a commonly adopted practice is to eliminate \emph{stop words} like ``a'', ``the'', \etc, from documents before indexing them for retrieval.
As for tweets, it is often the case that location-irrelevant words like ``downtown'' and ``OMG'' appear more frequently than ``howdy'' and ``phillies'' like words.
They will lead home location prediction results to random if indiscriminately taken into consideration.
Unlike eliminating predefined list of stop words, we usually resort to eliminate location-irrelevant words, \ie identify and keep local words.
Since local words are not enumerable like stop-words in most applications, a large amount of research efforts are spent on identifying local words, either unsupervised or supervised.

Unsupervised local word identification methods aim at statistical measures that are directly computable on the data and are indicative of a word's locality.
Laere \etal~\cite{DBLP:journals/tkde/LaereQSD14} propose two types of local word selection methods. One leverages Kernel Density Estimation~\cite{silverman1986density} which spatially smooth term occurrences, and the other is based on Ripley's K statistic~\cite{ripley2005spatial} which measures term's geographical deviation.
Inspired by Inverse Document Frequency (IDF) in information retrieval, Ren \etal~\cite{ren2012you} and Han \etal~\cite{DBLP:conf/coling/HanCB12} propose Inverse Location Frequency (ILF) and Inverse City Frequency (ICF), respectively, to measure the locality of words.
Their assumption is that local words should be distributed in fewer locations and have larger ILF and ICF values.
Besides IR-based measures, some studies also resort to measures that have information theoretic interpretations, \eg information gain and maximum entropy in~\cite{DBLP:conf/coling/HanCB12}, and K-L divergence in~\cite{yamaguchi2014online}.
Their assumption is that the distributions of local words should be more biased than ordinary ones. Noted that Yamaguchi \etal~\cite{yamaguchi2014online} deal with streaming tweets which could update users' home location according to newly posted tweets.
In~\cite{hecht2011tweets}, Hecht \etal propose a \textsc{Calgari} score for words, which is similar to information theory based measures.
Mahmud \etal~\cite{mahmud2012tweet} apply a series of heuristic rules to select local words.
Han \etal~\cite{DBLP:journals/jair/HanCB14} report a comparison of statistical-based, information theory-based and heuristic-based methods on local words selection.

On the other hand, supervised methods are also considered in a number of studies.
In~\cite{cheng2010you}, Cheng \etal view the problem of local word identification as a classification problem.
First, they fit the geographical distribution of each word with spatial variation model by Backstrom \etal~\cite{backstrom2008spatial}.
The spatial variation model assumes that each word has a geographical center, a center frequency $C$, and a dispersion ratio $\alpha$.
The probability of seeing the word at a location with distance $d$ to the center is proportional to $Cd^{-\alpha}$.
In simple words, this model specifies a one-peak distribution at the center with exponential decay.
After the model is fit, the parameters are used as word features.
Second, they manually labeled 19,178 words in a dictionary as either local or non-local.
Finally, they train a classification model and apply it to all other words in the tweet dataset.
Ryoo and Moon~\cite{ryoo2014inferring} apply the above method~\cite{backstrom2008spatial} to a Korean tweet dataset, and achieve satisfactory results.

\subsubsection*{Modeling Spatial Word Usage}
\label{sssec:modelLocalWords}

After identifying local words, the next problem is how to use them to predict users' home locations.
Most studies model this problem in a probabilistic manner.
Researchers propose probabilistic models to characterize the conditional distribution of users' home locations w.r.t.\ their tweets contents, then decompose and concretize the model to make predictions.

A representative probabilistic model is introduced in Cheng \etal~\cite{cheng2010you}.
The distribution of user $u$'s home location $l$ given her tweet contents $S(u)$ is decomposed as
\[
P(l|u)\propto \sum_{w\in S(u)}P(l|w)P(w)\text{.}
\]
Here only local word $w$ are considered, and $P(w)$ denotes the probability of $w$ over the entire corpus.
After the decomposition, major efforts are spent on estimating the location distribution $P(l|w)$ of word $w$, or \emph{spatial word usage}.
It is reported that estimating $P(l|w)$ directly from the corpus is inferior.
The reason is that some $w$ may be unobserved in less populated locations, which does not mean that the location is irrelevant to $w$.
To relieve this sparsity problem, smoothing techniques need to be involved.
A special type of spatial words is location names in tweets. Li \etal~\cite{li2012multiple} observe that the probability of tweeting venue names is location-based at some time, while it is also random at other time. Thus they make it a two level estimation. A Bernoulli distribution is adopted to estimate whether a location name is posted randomly or based on location, following which a multinomial distribution is used to estimate the probability of tweeting the venue name from each location.

In the same work~\cite{cheng2010you}, Cheng \etal propose several \emph{explicit} smoothing methods.
The first method, Laplace smoothing (or add-one smoothing), increase word $w$'s count in all locations by one before normalizing it to produce a distribution.
This method ensures that all locations get positive probabilities.
However, it does not involve the geographical information in $l$.
They further propose another two smoothing methods, namely, state-level smoothing and grid-based neighborhood smoothing.
In those methods, a fixed portion of per-state or per-cell word counts are evenly distributed to only locations in the same state or cell, instead of all locations on the map.
In~\cite{ren2012you}, Ren \etal also consider an explicit smoothing technique called circular-based neighborhood smoothing.
On the other hand, some parameterized spatial word usage models, once fitted, have \emph{implicit} smoothing effects.
In~\cite{cheng2010you}, Cheng \etal treat the fitted spatial variation model in~\cite{backstrom2008spatial} as a smoothed distribution.
In an extension work~\cite{cheng2013content}, Cheng \etal generalize the one-peak model~\cite{backstrom2008spatial} by wave-like smoothing to allow multi-peaks for words distributions.
In the influence-based social closeness models~\cite{li2012towards} (see Section~\ref{sssec:friendshipCloseness}), Li \etal treat friends followed and location names mentioned by users uniformly, and use Gaussian models to fit their geographical usage.
Instead, Chang \etal~\cite{chang2012phillies} use Gaussian mixture models to fit spatial word usage.
Their model also allows multi-peaks and is implicitly smoothed.

\subsubsection{Location-Centric Methods}
\label{sssec:locationcentric}

Word-centric methods characterize local words distributions and infer locations from them. Some other studies adopt different methods that give locations more centric roles.

A few studies adopt classification-based approaches to home location prediction.
They treat users' statistics about local words as features, and all candidate locations as classification labels.
Hecht \etal~\cite{hecht2011tweets} select top 10,000 words with highest \textsc{Calgari} scores as local words.
Users are then represented as 10,000-dimensional term frequency vectors, and fed into a multinomial Naive Bayes classifier for training and home location prediction.
Similarly, Rahimi \etal's~\cite{DBLP:conf/naacl/RahimiVCB15} apply logistic regression on users' TF-IDF vectors.
Instead of selecting local words as features, they subject to a sparse $ l_{1} $ regularization penalty~\cite{tibshirani1996regression}.
Similarly, Cha \etal~\cite{DBLP:conf/icwsm/ChaGK15} use sparse coding and dictionary learning techniques for word feature selection.
In~\cite{mahmud2012tweet}, Mahmud \etal adopt a hierarchical ensemble algorithm to train two-level classifier ensembles on the granularity of timezone-city or state-city.
In their extension work~\cite{mahmud2014home}, they also propose identifying and removing travelling people from training data to improve the performance of home location classifiers.
A person is considered travelling if any two of her tweets were sent from locations with distance above 100 miles.
Wing and Baldridge~\cite{wing2014hierarchical} also resort to hierarchical classification~\cite{silla2011survey}.
Instead of adopting administrative partitions directly, they use k-d tree to achieve adaptive grids in their hierarchy.
This leads to better granularity for populated regions, and avoids unnecessarily over-representing less populated areas.

There are also studies that adopt information-retrieval-based approaches to home location prediction.
They treat locations as pseudo-documents that consist of all tweets whose users live here.
Given the pseudo-document of a user whose home location is to be predicted, locations with the most similar pseudo-documents are retrieved as prediction results.
Specifically, Wing \etal~\cite{Wing:2011:SSD:2002472.2002593} adopt a grid representation of locations.
They estimate a language model~\cite{ponte1998language} for each grid with its pseudo-document.
Good-Turing smoothing~\cite{good1953population} is applied to smooth the probability of unseen words.
Kullback-Leibler divergence is adopted as the similarity measure between location documents and user documents.
In their subsequent work~\cite{wing2014hierarchical}, they resort to adaptive grids as in~\cite{roller2012supervised}.
When geo-coordinates need to be reported instead of grids, they find that reporting the centroid of user locations in the grid yields better precision than reporting mid-points of the grid.

Besides traditional methods, some recent works also explore deep learning models to tackle home location prediction. By extending their previous work~\cite{DBLP:conf/aclnut/MiuraTTO16}, Miura \etal~\cite{DBLP:conf/acl/MiuraTTO17} propose a more sophisticated model.
They order a user's messages chronologically and apply sequential model RNN to encode the content. In virtue of attention mechanism, a global message representation which addresses important information could be obtained. Similar process is also applied on context, \ie location description and timezone. The combination of the three representations is then fed to a softmax layer to predict home location.
Rahimi \etal~\cite{DBLP:conf/acl/RahimiCB17} apply a multilayer perceptron (MLP) with one hidden layer to classify users' home locations.
They adopt $l_2$ normalized bag-of-words representation of a given user's tweet contents as input. The output is a predefined discretized region generated by either a k-d tree or k-means.

\subsection{Inference Based on Twitter Network}
\label{ssec:friendshipNetwork3}

Besides posting tweets, other major activities that users involve in on Twitter are to establish following relationship and interact with friends.
Like their tweet contents, users' social relationships may reveal their home locations as well.
In Section~\ref{sssec:friendshipDistance}, we review some friendship-based methods, where friends are assumed to have smaller home location distances.
Moreover, it is also argued in studies that social-closeness, which is based on friendship, interactions, and other implicit signals, are more reliable for estimating home distances than sole friendship.
These studies are reviewed in Section~\ref{sssec:friendshipCloseness}.
Finally, when multiple users' home locations are unknown and to be predicted, their home locations are not independent because they are directly or indirectly interlinked through the Twitter network.
This dependency cannot be captured by \emph{local} inference methods that predict one home location at a time.
In Section~\ref{sssec:networkLocalGlobal}, we demonstrate how \emph{global} inference methods are applied in some studies.

\subsubsection{Friendship-Based Methods}
\label{sssec:friendshipDistance}

In social science, the assumption of \emph{homophily}~\cite{mcpherson2001birds} suggests that similar people make contacts at a higher rate than dissimilar ones.
Given the task of predicting home locations based on Twitter network, a quick intuition may be that one's home location is very likely to coincide with her friends' home locations.
In the preliminary model of~\cite{ren2012you}, Ren \etal assume that the higher proportion of a user's friends live at a location, the higher probability for the user to stay at the same location.
Davis \etal~\cite{davis2011inferring} employ a similar approach to that of~\cite{ren2012you}, except that they only consider mutual friendship.
Rodrigues \etal~\cite{DBLP:journals/ijon/RodriguesAPOM16} model home location prediction with the Potts model~\cite{DBLP:series/acvpr/978-1-84800-278-4}, which aims to maximize global home co-location between mutual friends.
One drawback of the above three approaches is that they do not use the coordinates of home locations of a user's friends.
Locations are treated as a discrete set of objects, while the distance between them is ignored.

One of the earliest attempts to model friendship and home location distance is made by Backstrom \etal~\cite{backstrom2010find}.
Although this study is conducted on Facebook, we include it in this survey because of its impacts on later Twitter-based studies.
The authors analyze a large number of Facebook users with known home locations and their friendships.
They try to fit the probability of two users being friends w.r.t.\ their home distance with the following curve
\begin{equation}\label{eq:facebookDist}
P(u_i, u_j \text{ are friends }|dist(u_i,u_j)=x)=a(b+x)^{-c}\text{,}
\end{equation}
and find that $c=1$ produces a good fit.
In other words, the probability of friendship is inversely proportional to home distance (with intercept $b$).
Based on this model, given friends of a user and their home locations, the most probable home location for the user could be found, by maximizing the probability of generating all seen friendship links.

The aforementioned three methods all depend user home proximity solely on \emph{direct} friendship.
In other words, they implicitly assume that friendship observed on an online social network implies real off-line friendship, and thus close home distance.
This may be far from true.
In~\cite{kong2014spot}, Kong \etal find that a pair of friends has 83\% of chance to live within 10 kilometers if their common friends account for more than half of their friends, respectively.
The chance decreases to 2.4\% if the common friend ratio is limited to 10\%.
This implies that rich \emph{indirect} friendships on Twitter may better indicate off-line friendship between two users, and thus their home location proximity.
As is also observed by Kossinets \etal~\cite{kossinets2006empirical}, if two users $a$ and $b$ have relationship with many third users, $a$ and $b$ may possibly have a relationship.
Inspired by this, Kong \etal improve the model in~\cite{backstrom2010find} by considering cosine similarity between two users' friend collections in Eq.~\ref{eq:facebookDist}.
Rout \etal~\cite{rout2013s} also relate the probability a user lives in a city to the distribution of indirect friendships between the user and her friends at the location.
Miura \etal~\cite{DBLP:conf/acl/MiuraTTO17} encode user friendship information into a neural network model. Different from the other works, they separate users in connected network and their corresponding cities, and assign them user embeddings and city embeddings respectively. An attention mechanism is applied on the addition of user and city embeddings to draw useful information on home location prediction.

\subsubsection{Social-Closeness-Based Methods}
\label{sssec:friendshipCloseness}

\begin{table*}
	\caption{Summary of studies on home location prediction. Works in bold are state-of-the-art methods based on the corresponding metrics and data. The same notations are used in the following tables.}
	\label{tbl:homeLocation}

	\begin{tabular}{lp{0.6in}p{0.9in}p{1in}p{1.7in}p{0.63in}p{0.9in}}
		\toprule
		Work & Input & Method & Dataset & Ground Truth & Granularity & Metrics  \\
		\midrule
		\textbf{\cite{ren2012you}} & Content, network & Hybrid & Data from~\textbf{\cite{cheng2010you,kwak2010twitter}} & Most frequent geo-tagged city, location profile & City, town & \textbf{MeanED, Acc@k, Acc}   \\
		
		\cite{DBLP:conf/coling/HanCB12} & Content & Word-centric & Data from~\cite{roller2012supervised}, geo-tagged tweets & Most frequent geo-tagged city & City & MedianED, Acc, Acc@d, MeanED  \\
		
		\cite{han2013stacking} & Content, context & Hybrid & Data from~\cite{DBLP:conf/coling/HanCB12}, geo-tagged tweets & Most frequent geo-tagged city & City & MedianED, Acc, Acc@k    \\
		
		\cite{yamaguchi2014online} & Content & Word-centric & Tweets & Location profile & Grid & Precision, Recall, MedianED   \\
		
		\cite{hecht2011tweets} & Content & Word-centric & Tweets & Location profile & Country, state & Acc   \\
		
		\cite{mahmud2012tweet} & Content, context & Word-centric & Geo-tagged tweets & The earliest geo-tagged city & City, state, time-zone & Recall, Acc@d  \\
		
		\cite{DBLP:journals/jair/HanCB14} & Content, context & Classification & Data from~\cite{DBLP:conf/coling/HanCB12,roller2012supervised}, tweets & The earliest geo-tagged city, most frequent geo-tagged city & Country, city & MedianED, Acc, Acc@d \\
		
		\cite{cheng2010you,cheng2013content} & Content & Word-centric & Geo-tagged tweets & Most frequent geo-tagged city & City & MeanED, Acc@k, Acc   \\
		
		\cite{ryoo2014inferring} & Content & Word-centric & Tweets & Median geo-tagged coordinates & Coordinates & MeanED  \\
		
		\cite{li2012multiple} & Content, network & Hybrid & Tweets & Location profile & City & Acc@k  \\
		
		\cite{li2012towards} & Content, network & Hybrid & Tweets & Location profile & City & Acc, MeanED   \\
		
		\cite{chang2012phillies} & Content & Word-centric & Data from~\cite{cheng2010you} & Most frequent geo-tagged city & City & Acc, MeanED  \\
		
		\cite{DBLP:conf/naacl/RahimiVCB15} & Content, network & Hybrid & Data from ~\cite{eisenstein2010latent,roller2012supervised,DBLP:conf/coling/HanCB12} & The earlist geo-tagged coordinates, coordinates of the most frequent geo-tagged city & Coordinates & Acc@d, MeanED, MedianED  \\
		
		\cite{DBLP:conf/icwsm/ChaGK15} & Content & Location-centric & CMU GeoText data & Geo-tag & Coordinates & MeanED, MedianED  \\
		
		\cite{mahmud2014home} & Content, context & Location-centric & Geo-tagged tweets & The ealiest geo-tagged city & City & Recall, Acc  \\
		
		\cite{wing2014hierarchical} & Content & Location-centric & Data from \cite{roller2012supervised,DBLP:conf/coling/HanCB12} & Coordinates of the earlist tweet, coordinates of the most frequent geo-tagged city & Grid & MeanED, Acc@d, MedianED   \\
		
		\cite{Wing:2011:SSD:2002472.2002593} & Content & Location-centric & Wikipedia, data from~\cite{eisenstein2010latent} & Geo-tag & Grid & MeanED, MedianED \\
		
		\cite{roller2012supervised} & Content & Location-centric & Data from~\cite{eisenstein2010latent}, geo-tagged tweets & The earlist geo-tagged coordinates & Grid & Acc@d, MeanED, MedianED   \\
		
		\textbf{\cite{DBLP:conf/acl/MiuraTTO17}} & Content, network, context & Hybrid NN & \textbf{Data from~\cite{roller2012supervised}} and \textbf{W-NUT} &　The earlist geo-tagged coordinates, majority vote of the closest city center & \textbf{City} & \textbf{MedianED, Acc, Acc@d, MeanED}  \\
		
		\textbf{\cite{DBLP:conf/acl/RahimiCB17}} & Content & MLP & Data from ~\textbf{\cite{eisenstein2010latent,roller2012supervised,DBLP:conf/coling/HanCB12}} & The earlist geo-tagged coordinates, coordinates of the most frequent geo-tagged city & \textbf{Grid} & \textbf{Acc@d, MeanED, MedianED} \\
		
		\cite{davis2011inferring} & Network & Friendship-only & Tweets & Most frequent geo-tagged city, location profile & City & Precision, Recall \\
		
		\cite{DBLP:journals/ijon/RodriguesAPOM16} & Content, network & Friendship-only & Geo-tagged tweets & Most frequent geo-tagged city & City & Precision, Recall, $F_{1}$, Acc  \\
		
		\cite{kong2014spot} & Network & Friendship-only & Tweets, Gowalla check-in & Most frequent check-in, location profile & Coordinates & Acc, MeanED  \\
		
		\cite{rout2013s} & Network & Friendship-only & Tweets & Location profile & City & Acc@d, MeanED   \\
		
		\cite{mcgee2013location} & Network & Social-closeness based & Geo-tagged tweets & Median geo-tagged coordinates & Coordinates & Acc@d, MeanED  \\
		
		\cite{compton2014geotagging} & Network & Social-closeness based & Geo-tagged tweets & Location profile, median geo-tagged coordinates & Coordinates & Recall, MeanED, MedianED \\
		
		\cite{jurgens2013s} & Network & Social-closeness based & Geo-tagged tweets, Foursquare data & Location profile, median geo-tagged coordinates & Coordinates & MedianED   \\
		
		\cite{yamaguchi2013landmark} & Network & Social-closeness based & Data from~\cite{li2012towards} & Location profile & Coordinates & Acc@d, Recall, $F_{1}$, MedianED, MeanED  \\
		
		\cite{DBLP:conf/acl/RahimiCB15} & Content, network & Hybrid & Data from \cite{eisenstein2010latent,roller2012supervised,DBLP:conf/coling/HanCB12} & The earlist geo-tagged coordinates, coordinates of the most frequent geo-tagged city & Coordinates & Acc@d, MeanED, MedianED \\
		
		\cite{eisenstein2010latent} & Content & Geo-topic & Geo-tagged tweets & The earliest geo-tagged city & State & MeanED, MedianED \\
				
		\cite{eisenstein2011sparse} & Content & Geo-topic & Data from~\cite{eisenstein2010latent} & The earliest geo-tagged city & Coordinates & MeanED, MedianED    \\
		
		\cite{DBLP:conf/asunam/EfstathiadesAPD15} & Context & Probabilistic & Data from~\cite{eisenstein2010latent}, geo-tagged tweets & Location profile, work place on LinkedIn & POI & MeanED, Acc@d, Acc  \\
		
		\cite{DBLP:conf/ht/PoulstonSB17} & Context & Clustering & Geo-tagged tweets & Manual label & Coordinates & MeanED, Acc \\
		
		\cite{DBLP:conf/emnlp/RahimiBC17} & Context & NN with mixture density network & Data from ~\cite{eisenstein2010latent,roller2012supervised} & The earlist geo-tagged coordinates & Coordinates & Acc@d, Meand ED, MedianED \\
		
		\bottomrule
	\end{tabular} 		

\end{table*}

In the previous subsection, we discussed several friendship-only methods, which only involve friendships available in the Twitter network.
However, it may harm home prediction if we depend home distance purely on direct friendship on Twitter.
Studies report that the inverse proportion model~\cite{backstrom2010find} in Eq.~\ref{eq:facebookDist} on Facebook does not hold for Twitter.
For example, McGee \etal~\cite{mcgee2011geographic} observe that friendship probability w.r.t\ home distance on Twitter roughly satisfy a bimodal distribution.
One peak is around 10 miles, and the other is far away.
Similar observations are also made by Scellato \etal~\cite{scellato2011socio} and Volkovich \etal~\cite{volkovich2012length} on other social networks. Investigations in~\cite{kong2014spot} and~\cite{rout2013s}  indicate that \emph{social closeness}, or how familiar two users are to each other in real life, is a better indicator of home proximity. Therefore, many subsequent works are dedicated to going beyond online friendship and estimating social closeness instead.

In Twitter network, \textit{mention} is another form of user interaction. When users mention each other or have conversation with each other, the two users are believed to have closer relationship or share similar interest. Such kind of `friendship' is valuable in home location prediction.
McGee \etal~\cite{mcgee2011geographic} make an analysis on 104,214 Twitter users with home located inside US.
They find that besides mutual friendship through following, users' actions of mentioning and actively chatting with each other also indicate their home proximity.
In a subsequent work~\cite{mcgee2013location}, McGee \etal confirmed similar observations by examining a larger dataset.
They also make more observations: 1) if the followed user account is a protected account\footnote{A protected account means that others need to get permissions to follow it, and its friend list and tweets are not public.} (typically an ordinary person), the two users are geographically close; and 2) local newspaper accounts are close to their followers.
By treating geographical proximity as ground truth social closeness, McGee \etal trained a decision tree to assign social closeness between different users to ten quantiles with the above cues as features.
They further use home distance in each social closeness quantile to fit Eq.~\ref{eq:facebookDist}, one model for each quantile.

Similar to McGee \etal~\cite{mcgee2011geographic,mcgee2013location}, Compton \etal~\cite{compton2014geotagging} also exploit mentions between users.
They build a user mention graph and optimize unknown home locations such that users mentioning each other are located as close as possible.
Jurgens~\cite{jurgens2013s} also considers bidirectional mention relationship instead of friendship. Rahimi \etal \cite{DBLP:conf/naacl/RahimiVCB15} find that bidirectional mention are too rare to be useful. They adopt unidirectional mention as undirected edge.

Besides mentions and conversations as social closeness indicators, some studies also suggest \emph{influence} to be another, but negative factor of social closeness.
For example, a user in Chicago may follow Lady Gaga in New York and President Obama in Washington.
The establishment of such following relationship is not a result of social closeness between the user and the celebrities, but caused by the celebrities' social influence.
The intuition in this example has been supported by a few studies.
By analyzing a large Twitter dataset, Kwak \etal~\cite{kwak2010twitter} find that users with fewer than 2,000 mutual friends (thus unlikely to have large influence) are more likely to be geographically close to most of them.
In McGee's work~\cite{mcgee2013location} described earlier in this subsection, they also discover that a user $u$'s friend who has many friends and followers tend to be further away from $u$.

In~\cite{li2012towards}, Li \etal construct a user influence model to capture the above intuitions.
Specifically, they model a user's influence as a bivariate Gaussian distribution centered at her location, with the variance of the distribution interpreted as her influence scope.
The probability of user $u_i$ following $u_j$ is measured by the probability density of $u_j$'s influence distribution at $u_i$'s home location.
Finally, all unknown home locations and influence scopes are treated as parameters and learnt from the data by Maximum Likelihood Estimation (MLE).
Similarly, Yamaguchi \etal~\cite{yamaguchi2013landmark} propose a \emph{landmark}-based home location prediction technique.
Here, a landmark is a user with a lot of friends living in a small region.
They argue that landmark friends are reliable cues to infer a user's home location.
In this sense, landmarks are actually non-celebrities with small influence.
In an extension~\cite{li2012multiple} of their earlier work~\cite{li2012towards}, Li \etal extend home location prediction to multiple location profiling.
The motivation is that many people may have home cities, as well as working and college cities that may not coincide with their homes.
They may not only follow friends living nearby and celebrities far away, but also colleagues and classmates in her working and college cities, respectively.

\subsubsection{Local vs. Global Inference}
\label{sssec:networkLocalGlobal}

Given that users are connected by the Twitter network, predicting their home locations is technically different from a typical prediction task where objects to be classified/scored are independent.
For most studies reviewed above, we only describe how to conduct \emph{local} inference, \ie predict a user's home location based on one- or two-hop friendship or mentioning.
Even if friendship-based and social-closeness-based features are carefully designed, one may still face many problems when implementing a home location predictor.
What if all friends of the current user have unknown home locations?
Whether and how should an inferred home location be updated when the user's friends' home locations are updated via inference?
In this subsection, we review some studies on how they deal with the above problems and how \emph{global} inference is carried out.

The easiest global inference approach would be to apply local inference \emph{iteratively} on users with unknown home locations (\ie label propagation~\cite{zhu2002learning}). In each iteration, a user's home is temporally guessed through their friends with known or inferred locations.
A few studies adopt this approach~\cite{backstrom2010find,jurgens2013s,DBLP:conf/naacl/RahimiVCB15,DBLP:conf/acl/RahimiCB15}.
However, it is also reported in~\cite{rout2013s} that simple iterations may reduce prediction accuracy.
The authors find that iteratively making prediction causes the population distribution to be flatter, which contradicts with the common sense that most people live in densely populated areas.
Therefore, they stick to local inference.
In~\cite{kong2014spot}, Kong \etal conduct a variation of iterative inference called \emph{confidence-based} iteration.
The idea is to estimate a confidence for each home location guess, and only pass those with high confidence to the next iteration.
Finally, it is worth noting that some studies define an explicit global objective function (or joint distribution) to optimize. Their inference methods are thus naturally global.
Rahimi \etal~\cite{DBLP:conf/acl/RahimiCB15} also find that  label propagation would be biased by highly-connected nodes (\ie celebrities with large amount of followers), and the nodes that are not connected to any labeled nodes could not be inferred. Therefore, they remove celebrities by identifying the number of mentions based on a graph constructed by mention relationship. For nodes with no labeled neighbors, they estimate the labels by the content-based method proposed in~\cite{DBLP:conf/naacl/RahimiVCB15}.
In~\cite{li2012towards}, Li \etal derive from their global likelihood a two-stage iterative maximization method.
Both unknown locations and influence scopes (recall in Section~\ref{sssec:friendshipCloseness}) are updated in each iteration.
Compton \etal~\cite{compton2014geotagging} directly optimize their objective function by parallel coordinate descent~\cite{richtarik2016parallel}.
On the other hand, Rodrigues \etal~\cite{DBLP:journals/ijon/RodriguesAPOM16} and Li \etal~\cite{li2012multiple} resort to Gibbs sampling~\cite{DBLP:journals/ml/AndrieuFDJ03} to infer parameters in their joint distributions.

\subsection{Inference based on Tweet Context}
\label{ssec:tweetContext3}

In Section~\ref{sssec:context}, we categorize various information associated with tweets as tweet context.
Among them, tweet posting time and self-declared user profiles like locations and time zones are mainly employed information to help predict home location.

Mahmud \etal~\cite{mahmud2012tweet,mahmud2014home} takes tweet posting time into consideration.
In their dataset, all posting times are recorded in GMT.
After binning a GMT day into time slots of equal length, users are viewed as distributions of tweet posting times.
Since users in different time zones exhibit time shifts in their distribution, a time-zone classifier is trained with the distribution as features.
Such classifications reveal the time zones of users and could provide a broad range of users' locations.
In the work of Han \etal~\cite{han2013stacking,DBLP:journals/jair/HanCB14}, the authors observe that self-declared locations and time zones, as free texts, are not always accurate. For example, informal abbreviations like ``mel'' (for Melbourne) may occur.
Therefore, besides tweet contents, they also include all four-grams of self-declared locations and time zones as features to train a home location classifier.
Efstathiades \etal~\cite{DBLP:conf/asunam/EfstathiadesAPD15} simply utilize a probabilistic model based on the temporal distribution of geo-tags associated with tweets to estimate user home location and work place. The method is based on their observation that tweeting activity during rest time (\ie late in the night) is more likely to be generated from ``home" location, while during working time posting activity is mostly likely to be generated from ``work" location.
Poulston \etal~\cite{DBLP:conf/ht/PoulstonSB17} also leverage geo-tags, but they find that users usually have several active regions. Simply adopting the median as home location is not appropriate. Thus they cluster the geo-tags first, and the group with highest number of posts is considered as ``home cluster". The geometric median of all points in ``home cluster" is taken as home coordinate.
Similarly, Cheng \etal~\cite{DBLP:conf/sigir/ChengCBB14} also group user's geo-tags into squares and the one with most number of geo-tags is regarded as the center. Instead of taking geometric median directly, they repeat the process within the center area with finer cells until the square size is smaller than a predefined size. The final center is considered as the user's home location.
By leveraging neural network model together with mixture density network, Rahimi \etal~\cite{DBLP:conf/emnlp/RahimiBC17} convert two-dimensional geo-tags into continuous vector space and take them as input.

\subsection{Summaries and Discussions}
\label{ssec:summaryHome}

In this section, we review literatures on user home location prediction.
We summarize the studies listed in Table~\ref{tbl:homeLocation}.
Techniques for home location prediction rely equally on tweet content and Twitter network.
For tweet content, word-centric approaches are characterized by two components, \ie local word identification and spatial word usage modeling.
Location-centric approaches, on the other hand, cast the problem to classification or ranking problems.
For Twitter network, dependencies between users' home locations are explained by their friendship and interactions.
Global inference approaches are involved to solve the collective inference problem.
Finally, tweet contexts like posting time and self-declared profiles are also involved in some studies.

Finally, we note that a systematic experimental comparison is conducted by Jurgens \etal~\cite{jurgens2015geolocation}.
The competing methods include Backstrom \etal~\cite{backstrom2010find}, Kong \etal~\cite{kong2014spot}, Li \etal~\cite{li2012towards,li2012multiple}, Mcgee \etal~\cite{mcgee2013location}, Rout \etal~\cite{rout2013s}, Davis \etal~\cite{davis2011inferring}, Jurgens~\cite{jurgens2013s} and Compton \etal~\cite{compton2014geotagging}.
Their dataset consists of 1.3 billion tweets, 15 million users, and 26 million following relationships.
Both self-declared home location and aggregation of geo-tags have been adopted as ground truth.
Readers can refer to this experimental comparison for detailed results.

\section{Tweet Location Prediction}
\label{sec:tweetLocation}

According to an analysis by Java \etal~\cite{java2009we}, users' primary aims of sending out tweets are to share or to seek information.
For example, one may tweet about a restaurant where she is enjoying delicious food.
Such information will help promote the restaurant, if its name is clearly associated with the tweet as a tag.
One may also send tweets saying she is lost when looking for a building.
In this case, a tag on where the tweet is posted may enable her friends to give her precise directions.
Unfortunately, it is reported that less than 1\% of tweets have explicit geo-tags~\cite{graham2014world}.
Therefore, predicting tweet location has received considerable attention.

At the first glance, tweet location prediction seems to be very similar to home location prediction.
The ``only'' difference seems to be their inputs: for home location prediction we have all tweets from a user, while for tweet location prediction we are given only one tweet.
In this section, we review literatures on tweet location prediction.
We will also spend efforts to highlight different properties of the two problems, as well as different emphasis resulted on specific techniques.

\subsection{Inference based on Tweet Content}
\label{ssec:tweetContent4}

Due to similar problem definitions, tweet location and home location predictions share many common techniques on handling tweet content.
For example, word-centric and location-centric methods, which we reviewed for home location prediction, are also observed in studies on tweet location prediction.
We will detail those works in Section~\ref{sssec:wordLocationCentric}.
Moreover, we will also review some topic-model-based approaches in Section~\ref{sssec:geo-topic}, which are (most of the time) specially designed for tweet location prediction.

\subsubsection{Word- or Location-Centric Methods}
\label{sssec:wordLocationCentric}

As summarized in Section~\ref{ssec:tweetContent3}, word-centric methods for home location prediction~\cite{li2012multiple,li2012towards,chang2012phillies} are characterized by modeling spatial word usage.
To alleviate the data sparsity issue, Gaussian or Gaussian mixture models are used to achieve smoothed word usage distributions~\cite{li2012multiple,li2012towards,chang2012phillies}.
Similarly, in~\cite{priedhorsky2014inferring}, Priedhorsky \etal also employ Gaussian mixture models for tweet location prediction.
However, they concentrate on modeling the spatial usage of not only words, but also n-grams.
The reason lies in that, for tweet location prediction, we have only one tweet as the input.
This information is much more limited than that for home location prediction, where a large number of tweets from a user are provided.
Therefore, it is worthwhile to exploit the input with reasonable redundancy.
In experiments, they find that their models are improved by including rare n-grams, even those occurring just three times.
Flatow \etal~\cite{flatow2015accuracy} also resort to modeling spatial n-gram usage with Gaussian models.
Similar to the idea of local words, they prefer \emph{geo-specific} n-grams, \ie those whose tweets are mostly located in a small eclipse on the map.
Alternatively, Chong and Lim~\cite{DBLP:conf/icwsm/ChongL17} apply a learning to rank method which encodes tweet content by a smoothed probability estimation that a word occurs at a venue. In their following work~\cite{DBLP:conf/cikm/ChongL17}, word importance for different locations is distinguished.
Since single tweet is short and of little information, they borrow the idea of query expansion and add words from the user's related historical tweets as supplement information. This is based on the assumption that users tend to visit same or related locations because of habits or constrains.

As for location-centric methods, previous studies also involve information-retrieval-based solutions.
Kinsella \etal~\cite{kinsella2011m} treat both tweets and locations as Dirichlet-smoothed~\cite{Zhai:2001:SSM:383952.384019} unigram language models.
The probability of a location language model generating a tweet, or the KL-divergence between language models of a tweet and a location, are adopted as location ranking functions.
Li \etal~\cite{Li2011the} also employ an information-retrieval-based approach with KL-divergence as the retrieval function.
For locations with few tweets, they augment their language models with web pages retrieved through their names.
Similarly, Lee \etal~\cite{DBLP:conf/mobiquitous/LeeGSL14} resort to user tips posted on the Foursquare pages of locations to construct language models for those locations.
Besides Laplace smoothing (or add-one smoothing), they also try absolute discounting and Jelinek-Mercer smoothing~\cite{DBLP:conf/acl/ChenG96} to deal with unseen words, but no performance gain is observed.
Liu and Huang~\cite{DBLP:conf/cikm/Liu016} apply Hidden-Markov-based model to infer tweet location on city-level. The observations are language models for each city based on geo-tagged tweets and the states are corresponding cities.

We also note that a few tweet location prediction studies involve classification-based approaches.
Hulden \etal~\cite{DBLP:conf/aaai/HuldenSF15} classify tweet text into discretized cell grids with words as features. A data sparsity issue appears when grid size becomes too small. To deal with this problem, they apply a Gaussian kernel to estimate the prior probability of each cell and the conditional probability of each word given a cell.
Besides unigrams, Dredze \etal~\cite{DBLP:conf/naacl/DredzeOK16} also extract bigrams from tweet content, together with features derived from Twitter contexts, and feed them to a classifier.
Cao \etal~\cite{DBLP:conf/bigdataconf/CaoCJY15} employ both tweet content and social relationship features to classify tweet text to locations at fine-grained POI level.
Another work~\cite{hahmann2014twitter} we are aware of aims at predicting location types, \eg railway station, cinema or supermarket, rather than exact locations for tweets.
The underlying reason may be again due to the large number of fine-grained tweet locations.
For user home prediction, the number of classes, \ie cities, are manageable under the multi-class classification framework.
Some works even alleviate the class number issue by hierarchical classification~\cite{mahmud2012tweet,mahmud2014home,wing2014hierarchical}.
However, the class number is simply unaffordable for tweet location prediction, given that there may be hundreds of thousands of POIs in a city.
Iso \etal~\cite{DBLP:journals/corr/IsoWA17} adopt Neural Network model to predict tweet location. They utilize convolutional mixture density network which is fed by tweet content, to estimate the parameters of Gaussian mixture model, and employ the mode value of estimated density as the predicted coordinates for tweets. They claim that different loss functions do affect  model performance.

\begin{table*}
	\caption{Summary of studies on tweet location prediction.}
	\label{tbl:tweetLocation}
	\begin{tabular}{lp{0.45in}p{1.2in}p{1.1in}p{1.15in}p{0.9in}p{0.9in}}		
		\toprule
		
		Work & Input & Model & Dataset & Ground Truth & Granularity & Metrics   \\
		\midrule
		\cite{priedhorsky2014inferring} & Content & Word-centric & Data from \cite{eisenstein2010latent}, geo-tagged tweets & Geo-tag & Coordinates & MeanED, Precision, Recall  \\
		
		\textbf{\cite{flatow2015accuracy}} & Content & Word-centric & \textbf{Geo-tagged tweets} & Geo-tag & Coordinates & \textbf{MeanED, Precision, Recall,} $ \mathbf{F_{1}} $   \\
		
		\cite{DBLP:conf/icwsm/ChongL17} & Content, Context & Ranking & Foursquare data, tweets & Foursquare check-ins & POI & MRR\footnote{MMR} \\  
		
		\cite{DBLP:conf/cikm/ChongL17} & Content, network & Naive Bayes model & Foursquare data, tweets & Foursquare check-ins & POI & MRR, VMMR\footnote{MMR} \\  
		
		\cite{kinsella2011m} & Content & Location-centric & Geo-tagged tweets & Geo-tag & Country, state, city, zip-code & Acc, Acc@k  \\
		
		\cite{Li2011the} & Content, context & Location-centric & Geo-tagged tweets & Geo-tag & POI & Acc@k  \\
		
		\cite{DBLP:conf/mobiquitous/LeeGSL14} & Content & Location-centric & Foursquare data, geo-tagged tweets & Geo-tag & POI & Precision, Recall \\
		
		\cite{DBLP:conf/cikm/Liu016} & Content & Location-centric & Geo-tagged tweets & Geo-tag & City & MeanED, MedianED, Acc \\
		
		\cite{DBLP:conf/aaai/HuldenSF15} & Content & Classification & Data from~\cite{eisenstein2010latent}, geo-tagged tweets & The earliest geo-tagged coordinates, geo-tag & Coordinates & MeanED, MedianED \\
		
		\cite{DBLP:conf/naacl/DredzeOK16} & Content, context & Classification & Geo-tagged tweets & Geo-tag & Country, city & Acc, Acc@d, MedianED \\
		
		\cite{DBLP:conf/bigdataconf/CaoCJY15} & Content, network & Classification & Geo-tagged tweets, Foursquare data & Geo-tag & POI & Acc@k, MeanED \\
		
		\cite{hahmann2014twitter} & Content & Classification & Geo-tagged tweets & Human label & POI & Precision, Recall, Acc   \\   	
		
		\cite{DBLP:journals/corr/IsoWA17} & Content & Convolutional Mixture Density Network & Geo-tagged tweets & Geo-tag & Coordinates & MeanED, MedianED \\
		
		\cite{hong2012discovering} & Content & Geo-topic & Geo-tagged tweets & Geo-tag & Coordinates & MeanED  \\
		
		\cite{chen2013interest} & Content & Geo-topic & Geo-tagged Weibo data & Human label & POI & Acc, MeanED  \\
		
		\textbf{\cite{yuan2013and}} & Content, context & Geo-topic & Data from~\textbf{\cite{eisenstein2010latent}}, geo-tagged tweets & Geo-tag & Coordinates & \textbf{Acc, MeanED}  \\
		
		\cite{sadilek2012finding} & Network & Dynamic Bayesian network & Geo-tagged tweets & Geo-tag & Coordinates & Acc@d  \\
		
		\cite{schulz2013multi} & Content, context & Stacking & Geo-tagged tweets & Geo-tag & Coordinates & MSE, MedianED, MeanED, Recall  \\
		
		\cite{DBLP:conf/icaart/GalalE16} & Content, context & Classification & Geo-tagged tweets & Location check-ins & Location category & Acc  \\
				
		\bottomrule
	\end{tabular} 	
\end{table*}

\subsubsection{Geo-Topic-Model-Based Methods}
\label{sssec:geo-topic}

As effective approaches to unsupervised text mining, topic models have been extended to account for texts with geographical information like blogs~\cite{wang2007mining,mei2006probabilistic}.
Such models are also expanded to tweets and used for geolocation on tweets due to their generative nature. Topic models could integrate different aspects related to locations as latent variables into a unified model, which could make information interact with each other, as we call them geo-topic-model-based method.

Eisenstein \etal~\cite{eisenstein2010latent} extend traditional topic models by ``corrupting'' conventional topics and produce location-varied topics.
For example, ``NBA'' and ``Kobe'' may be representative words in ``basketball'' topic produced by conventional models.
By sampling from a Gaussian distribution centered at the ``basketball'' topic vector, the corrupted ``basketball'' topic for Boston may also include ``Celtics'' (a Boston-based team) while slightly changing other word frequencies.
In their subsequent work, Eisenstein \etal\cite{eisenstein2011sparse} propose a Sparse Additive GEnerative model (SAGE).
The model is capable of supporting the idea of location-based topic corruption in~\cite{eisenstein2010latent}.
It also enables sparsity and simplicity in model inference.
An issue in these works~\cite{eisenstein2010latent,eisenstein2011sparse} lies in the special way they pre-process tweets.
They concatenate each user's tweets into a long tweet, and use the first valid geographical coordinates as the location of the long tweet.
We note that the two works are actually for home location prediction, and introduced here for the sake of a complete review of topic-model-based methods.

By leveraging SAGE model~\cite{eisenstein2011sparse}, Hong \etal~\cite{hong2012discovering} construct a model that takes region, topic and users' interests into consideration.
Different from~\cite{eisenstein2010latent,eisenstein2011sparse}, they respect the original view of tweets and model locations in a per-tweet manner.
They assume tweet location depends on the user's geographical interest distribution.
The topic of a tweet then depends on the user's topical interest, as well as local topics.
Words in the tweet are finally generated by the chosen topic as well as a ``local words'' distribution.
Instead of modeling users' geographical interest as a multinomial distribution, Chen \etal~\cite{chen2013interest} introduce user interest as a latent variable and construct a \emph{location function}, \eg eating, shopping, or health, all of which are as bridges to link users and locations.
Each user has an interest distribution over location functions, which affect tweets generation.
Yuan \etal~\cite{yuan2013and} propose an intermediate variable called \emph{regions} between users and tweet locations.
For example, a user may have a ``work'' region and a ``home'' region, which are Gaussian distributions centered at her work place and home address, respectively.
Suppose the user is at her work region and wants to eat, \ie choosing ``eating'' from her topical interests.
She will pick a restaurant near her work place and write a tweet about eating and the work region, tagged with the name of the restaurant.

\subsection{Inference based on Twitter Network}
\label{ssec:friendship4}

Compared with home locations, tweet locations are usually described at a much finer granularity, \ie POI-level rather than city-level, and are highly dynamic. Besides, tweets are usually short and noisy which increase the difficulty of predicting tweet location. To enrich available information, some works also try to align with friendship network.

In Sadilek's work~\cite{sadilek2012finding}, the dynamic input comes from real-time locations of a user's friends, and her own historical locations.
To study the correlation between the trajectories of friends and the auto-correlation within one's trajectory, they accumulate over ten thousands of users, each with more than one hundred geo-tagged tweets.
A Dynamic Bayesian Network (DBN) is trained on the location sequence of each user, with her friends' locations, the time of the day, and the day of the week as features.
One interesting aspect of their model is that it not only models the attractive force between friends' locations but also captures other non-linear patterns.
For example, two co-workers in the same store may have a day shift and a night shift.
In this case, given enough historical data, their model can predict that one is at home given that the other is working in store.
Chong and Lim~\cite{DBLP:conf/cikm/ChongL17} find that users with more similar tweet content history may be more similar in their venue visitation history. Collaborative filtering is adopted to propagate visitation information to users without location visiting history based on the similarity of historical tweet content. They provide us a new view that useful information can be obtained even from users without following or followed relationship.

\subsection{Inference based on Tweet Contexts}
\label{ssec:tweetContext4}

Tweet posting times are indicative of users' home locations, where a user is characterized by a distribution of posting times~\cite{mahmud2012tweet,mahmud2014home}.
Unlike home locations, for tweet location prediction we only access a tweet's posting time rather than a distribution.
However, a time stamp may also be informative if enough historical data for locations are provided.
For example, tweet posting histories may suggest that a club tends to be tweet-active at night, while a park tends to receive more tweets on weekends.
Inspired by this, Li \etal~\cite{Li2011the} keep tweet time distributions for locations at three different scales of periods, \ie day, week, and month.
Given a tweet with a timestamp, probabilities of the three distributions generating the timestamp are linearly combined to give preferences between locations.
In the geographic topic model of~\cite{yuan2013and}, Yuan \etal adopt two scales of time periods, namely day (weekday/weekend) and time of the day.
Given a user, the generative model first decides whether on weekdays or weekends to send the tweet according to her preference.
Then the daytime is drawn from her preference distribution, which is also conditioned on the day variable.
Finally, the user decides which region to go to and send a tweet about.
Dredze \etal~\cite{DBLP:conf/naacl/DredzeOK16} take both time zone and tweet posting time as features for a classifier. They find the cyclical temporal patterns do have effects on prediction results.

\footnotetext[8]{Mean Reciprocal Rank}
\footnotetext[9]{Macro-averaged version of Mean Reciprocal Rank}

Schulz \etal~\cite{schulz2013multi}, on the other hand, accumulate tweet location indicators from user profiles.
Possible indicators may be users' self-declared home locations, websites, and timezones, as well as location names mentioned in the tweet.
By querying multiple databases\footnote{Those include GADM database of Global Administrative Areas (\url{http://www.gadm.org}), ThematicMapping (\url{http://thematicmapping.org/downloads/world borders.php}), and IANA Time Zone Database (\url{http://efele.net/maps/tz/world/}).}, those indicators are resolved to polygon-shaped administrative regions, with resolution confidences being heights of the polygons.
Those polygons are finally stacked up~\cite{woodruff1994gipsy} to produce a spatial distribution of possible tweet locations. In experiments, they find that such a multi-indicator approach is more robust than single-indicator approaches, which is error-prone due to ambiguity.
Chong and Lim~\cite{DBLP:conf/icwsm/ChongL17} provide another angle to utilize the context information and observe that both venues' active time and users' visiting place histories could help on tweet location prediction.
They investigate venues' active time and estimate the probability that a location is popular given a time by a smoothed kernel density estimation method. Besides, they find an average user is spatially focused because she is usually constrained by geographical, social or personal factors. Thus, they encode this idea into the estimation of the probability that a user visits a location.

\subsection{Summaries and Discussions}
\label{ssec:summaryTweet}

As listed in Table~\ref{tbl:tweetLocation}, we review literatures on tweet location prediction.
Besides the fact that techniques for both tweet location and home location predictions emphasize much on employing tweet content, we also discuss several differences between home location and tweet location predictions.
We list them below for a concise summary:
\begin{itemize}
  \item Except studies with distance-based evaluations, home locations are predicted at coarse granularities like city, while tweet locations at a finer POI-level.
  \item Home location prediction relies equally on Twitter network and tweet content; but few studies utilize Twitter network to predict tweet locations.
  \item Classification-based approaches are common for home location prediction, which is not the case for tweet location prediction.
  \item When employing posting time information, users are viewed as time distributions, while tweets are essentially time stamps. This may leads to different location ranking functions.
\end{itemize}
Despite the above differences, we note that the two problems not always clearly separated.
Studies like~\cite{wing2014hierarchical,DBLP:conf/icwsm/ChaGK15,roller2012supervised,eisenstein2010latent,eisenstein2011sparse} concatenate a users' tweets into one document, and use the first available geo-tag as the ground-truth location.
We note that a geo-tag chosen this way may not necessarily to be the user's home location.
Since users are not explicitly modeled, their techniques could be used for both  prediction tasks.
On the other hand, \cite{hong2012discovering,chen2013interest,yuan2013and} explicitly model users' interests over locations, location functions, and regions.
These models may only be used for tweet location prediction, but better exploitation for the specific problem and data could be expected from them.

\section{Mentioned Location Prediction}
\label{sec:locationLinking}

Users occasionally send tweets to comment on a restaurant, a shopping mall, or a cinema, by treating Twitter as a life-logging platform.
When parades or disasters take place, numerous tweets may be sent out by users to inform others about the events.
Besides attaching geo-tags to those tweets, users may also reveal the relevant locations by mentioning their names in tweets.
Preprocessing on the location names are crucial steps to accumulating information for, and performing subsequent analysis on, users and events~\cite{imran2014processing,maceachren2011senseplace2}.
There are two steps for mentioned location processing: 1) recognition: to label text chunks which are potential location mentions, and 2) disambiguation: to map recognized location mentions to the right entries in a location database.

For well-formatted documents (\eg news), the entity recognition~\cite{nadeau2007survey} and disambiguation~\cite{DBLP:journals/tkde/ShenWH15} problems have been investigated for decades.
It is well received that the \emph{variability} and \emph{ambiguity} of entity mentions are two major difficulties for entity recognition and linking.
Here variability means an entity may be mentioned in various surface forms, and ambiguity means one mention may refers to multiple entities.
Unfortunately, the two difficulties are actually rendered more challenging by the noisy and short nature of tweets.
In this section, we review recognition and disambiguation efforts for location mentions in tweets.
We highlight how the two problems are made worse in the tweet scenario, and how they are dealt with by existing studies.
Note that, we may not limit in studies solely on location entities.
Recognition and disambiguation efforts of other types of entities in tweets will also be included in our survey, as long as they are inspiring to, and experimentally involve, mentioned location prediction.

\subsection{Inference based on Tweet Content}
\label{ssec:tweetContent5}

Like in ordinary documents, recognizing and disambiguating mentioned locations in tweets are generally based on tweet content, and are carried out in a pipelined manner.
On the one hand, words like ``Street'' and ``at'' may suggest inner and outer boundaries of location mentions.
On the other hand, other words in the mention context may provide clues for disambiguating the mentions.
We will introduce previous works on both tasks in Sections~\ref{sssec:NER} and~\ref{sssec:linking}, respectively.
We also note that some studies propose joint approaches to couple the two tasks.
They will be reviewed in the end of Section~\ref{sssec:linking}.

\begin{table*}
	\centering
	\caption{Summary of studies on named entity recognition. The ground truth in these studies are all based on human annotation.}
	\label{tbl:ner}
	\begin{tabular}{p{0.6in}p{1.4in}p{1.6in}p{0.85in}p{0.8in}p{1in}}  %
		\toprule
		Work & Input & Method & Dataset & NER type & Metrics    \\
		\midrule
		
		\cite{ritter2011named} & POS tagging, shallow parsing, capitalization & CRF & Tweets, Freebase & Location, person, \etc  & Precision, Recall, $F_{1}$ \\  
		
		\cite{liu2011recognizing,liu2013named} & Contextual, dictionary, orthographic, lexical  & KNN, CRF & Tweets & Location, person, \etc & Precision, Recall, $F_{1}$  \\
		
		\cite{li2012twiner} & Dictionary, statistical & Dynamic programming & Microsoft Web N-Gram, tweets & Location, person, \etc & Precision, Recall, $F_{1}$   \\
		
		\cite{malmasi2015location} & POS tagging & Rule-based matching & Tweets & Location & Precision, Recall, $F_{1}$  \\
		
		\cite{gelernter2013cross} & Lemma form, POS tagging, capitalization, dictionary, contextual, orthographic & Named location recognizer, street and building parser, NER & Tweets, GeoNames & Location & Precision, Recall, $F_{1}$   \\
		
		\cite{gelernter2013algorithm} & Orthographic & NER, gazetteer matching, lexico-semantic pattern recognition & NGA gazetteer, tweets  & Location & Precision, Recall, $F_{1}$  \\
		
		\textbf{\cite{li2014fine,liextracting}} & Lexical, contextual, grammatical, BILOU schema, geographical & CRF & \textbf{Tweets, Foursquare} & \textbf{Location} & \textbf{Precision, Recall,} $\mathbf{F_{1}} $  \\
		
		\bottomrule
	\end{tabular} 	
\end{table*}

\begin{table*}
	\centering
	\caption{Summary of models on location mention linking. All the studies are on the granularity of POI level.}
	\label{tbl:locationMention}
	\begin{tabular}{llp{1.4in}p{1.45in}ll}
		\toprule
		Work & Input & Model & Dataset & Ground Truth & Metrics   \\
		\midrule
		\cite{zhang2015geocoding} & Content & Classification & Geo-tagged tweets & Human label & Precision, Recall  \\
		
		\textbf{\cite{DBLP:conf/www/JiSCH16}} & Content & Structured perceptron with multi-view learning & \textbf{Tweets} & Human label & \textbf{Precision, Recall,} $\mathbf{F_{1}}$   \\
		
		\cite{li2014effective} & Content & Ranking & Foursquare data, Geo-tagged tweets & Geo-tag & Precision, Recall, $F_{1}$  \\
		
		\cite{DBLP:conf/kdd/ShenWLW13} & Content & Graph-based & Tweets & Human label & Acc \\
		
		\cite{DBLP:conf/naacl/GuoCK13} & Content & Structural SVM & Tweets, some data from~\cite{ritter2011named} & Human label & Precision, Recall, $F_{1}$   \\
		
		\textbf{\cite{DBLP:conf/sigmod/HuaZZ15}} & Network, context & Ranking & \textbf{Tweets} & NER identified by~\cite{li2012twiner} & \textbf{Acc} \\
		
		\cite{DBLP:journals/tacl/FangC14} & Content, context & Probabilistic & Geo-tagged tweets & Human label & Precision, Recall, $F_{1}$   \\
		
		\bottomrule
	\end{tabular} 	
\end{table*}

\subsubsection{Mentioned Location Recognition}
\label{sssec:NER}

For Named entity recognition (NER) in formal documents, state-of-the-art machine learning algorithms like conditional random fields~\cite{lafferty2001conditional} have been designed.
Equipped with comprehensive linguistic features like Part-of-Speech (POS) tags and capitalizations, they could achieve satisfactory performance~\cite{ratinov2009design}.
Based on those algorithms and features, off-the-shelf NER tools like \textit{StanfordNER}\footnote{http://nlp.stanford.edu/software/CRF-NER.shtml} and \textit{OpenNLP}\footnote{http://opennlp.apache.org} are also developed and released.

When faced with noisy and short tweets, traditional NER features and tools are both at risk of deteriorated performance.
For example, consider a typical tweet saying ``shopping @ orchard st''.
Because of the informal writing, common clues indicating ``Orchard Street'' as a location mention in formal documents, like ``at'' (``@''), ``street'' (``st''), and capitalizations (``Orchard'' instead of ``orchard''), are all absent.
Ritter \etal~\cite{ritter2011named} rebuild the entire NER pipeline for tweets.
They use Brown clustering~\cite{brown1992class} to identify word variations clusters (\eg ``at'' and ``@'').
A dedicated classifier is also trained to recognize whether each capitalization in a tweet is informative.
Similarly, Liu \etal~\cite{liu2013named,liu2011recognizing} train a tweet normalization model to correct informal words (\eg ``gooood'' to ``good") before performing NER.
Noticing that words like ``orchard'' may be hard to label within the given short tweet, they train a k-nearest-neighbor word classifier to inform the NER classifier with global information, \ie how the word is labeled in other tweets.
Li \etal~\cite{li2012twiner} investigate a novel streaming setting for tweet NER.
They exploit the gregarious property of entity mentions to differentiate valid mentions from non-entity segments.
Their approach also inherently addresses the short tweet problem.

Besides the above tweet NER attempts for general entities, there are also a few studies specially on location entities.
Those studies are characterized by the use of  location gazetteers, \eg \textit{Geonames}~\footnote{http://www.geonames.org/}~\cite{malmasi2015location,zhang2015geocoding} and \textit{Foursquare}~\cite{li2014fine,liextracting}.
Malmasi \etal~\cite{malmasi2015location} do not involve CRF in their location mention recognizer.
They simply use an off-the-shelf dependency parser to exact all noun phrases, and conduct fuzzy matching with \textit{Geonames}.
Their matching criteria take patterns of addresses and POIs into consideration.
Zhang \etal~\cite{zhang2015geocoding} rely on a location mention recognizer they build in~\cite{gelernter2013cross}.
A gazetteer-based location parser, a CRF-based recognizer, and a rule-based street/building parser are used in conjunction to achieve best recall.
A similar combination is also adopted by Gelernter \etal~\cite{gelernter2013algorithm}.
Li \etal~\cite{li2014fine,liextracting} observe that Twitter users often mention locations by abbreviations~\cite{lieberman2010geotagging}.
They opt to augment their Foursquare-based gazetteer with frequent-substring-based partial names.

\subsubsection{Mentioned Location Disambiguation}
\label{sssec:linking}

Given location mentions recognized in a document, location disambiguation (\ie linking)~\cite{DBLP:journals/tkde/ShenWH15} refers to resolving those mentions to right entries in a location database.
The challenge of this task lies in that different locations may have the same names.
For example, at the coarse city-level granularity, ``Washington'' may refer to a state in the west of the U.S., as well as a city in the east.
``Olympia'' may refer to the capital city of Washington state, as well as an ancient Greek city.
At a finer POI-level, chained restaurants, \eg McDonald, may have many branches  in a city.

For general entities in formal documents, traditional approaches~\cite{dill2003semtag,mihalcea2007wikify,milne2008learning} disambiguate one mention at a time.
To exploit dependencies between mentions, pair-wise fashioned~\cite{witten2008effective,kulkarni2009collective} and global collective disambiguation approaches~\cite{cucerzan2007large,hoffart2011robust,DBLP:conf/sigir/HanSZ11} are proposed.
Those approaches assume that the disambiguation decisions for multiple mentions in the same document should be \emph{coherent}.
For example, if ``Washington'' and ``Olympia'' co-occur in the same tweet, they are more likely to refer to the U.S. state and its capital.
As for mentioned locations in tweets, Zhang \etal~\cite{zhang2015geocoding} employ similar ideas in their study. They take the hierarchy structure of locations into consideration. Not only parent-child location pairs (\eg ``Washington" and ``Olympia"), but also siblings in the location hierarchy (\eg cities in the same state), are regarded as coherent.
Ji \etal~\cite{DBLP:conf/www/JiSCH16} investigate collectively disambiguating POI mentions in tweets.
Their coherence measure is based on the average distance among chosen POIs for the recognized mentions.
Different from~\cite{zhang2015geocoding,DBLP:conf/www/JiSCH16}, Li \etal~\cite{li2014effective} advocate disambiguation coherence at user-level rather than tweet-level.
They assume that mentioned locations in a user's tweets are generally inside her living city.
They first identify the living city by aggregating candidate locations for the mentions, and then refine those candidates with the living city.
Shen \etal~\cite{DBLP:conf/kdd/ShenWLW13} also conduct collective disambiguation at user-level by modeling user interests.
However, their method is aimed for general entities.

In conventional studies, mentioned location disambiguation is based on the output of recognition in a pipeline manner.
If fed with wrong outputs, \eg mentions with inaccurate boundaries, the disambiguation component may fail due to inability of finding candidates in the database.
Motivated by this, recent studies~\cite{DBLP:conf/naacl/GuoCK13,DBLP:conf/www/JiSCH16} suggest enabling information to flow in both directions between the two components.
If the disambiguation component suffers from no candidates or low confidence, it may give feedbacks to the recognition component to correct the input mentions.
In~\cite{DBLP:conf/naacl/GuoCK13}, Guo \etal leverage structural SVMs~\cite{tsochantaridis2005large} to jointly optimize mention recognition and disambiguation.
Both recognition features (\eg capitalization) and disambiguation features (\eg entity popularity) are integrated to train the structural SVM.
Similarly, Ji \etal~\cite{DBLP:conf/www/JiSCH16} jointly consider both types of features in a structural prediction framework.
They resort to beam search~\cite{DBLP:conf/acl/ZhangC08} to look for the best combination of recognition and disambiguation decisions.

\subsection{Inference Based on Twitter Network, Tweet Context}
\label{ssec:friendshipNetwork5}

Like home and tweet location prediction, user friendship and contextual information could also be explored for mention disambiguation.

In~\cite{DBLP:conf/sigmod/HuaZZ15}, Hua \etal assume that the more a user is influenced by others mentioning an entity, the more likely she will mention the same entity.
Specifically, they adopt an incremental disambiguation approach.
In the offline stage, they preprocess a large number of tweets with~\cite{DBLP:conf/kdd/ShenWLW13} as a base system.
Such preprocess enables them to estimate friendship-based user interest for entities in the online stage.
When a candidate entity $e$ is considered for a mention in user $u$'s tweet, they look for other users who once mentioned $e$. An entity $e$ is preferred if its users have good reachability to $u$ in the friendship network.
Besides friendship network, they also exploit time stamps of tweets. Due to their incremental disambiguation framework, they could estimate \emph{entity recency} when a new tweet comes. Given a time stamp, the recency for an entity $e$ is defined by the number of tweets mentioning $e$ in the last time window of predefined length. They further use personalized PageRank~\cite{jeh2003scaling} to propagate entity recency on the Wikipedia network to account for related entities. Finally, recently hot entities are rewarded when disambiguating mentions.

Fang \etal~\cite{DBLP:journals/tacl/FangC14} consider both geo-tags and time stamps of tweets in mention disambiguation.
An entity prior w.r.t. time and location is estimated and used to replace the coarse-grained global entity popularity.
Note that~\cite{DBLP:conf/sigmod/HuaZZ15,DBLP:journals/tacl/FangC14} aim for general entities, not limiting to locations.
When only locations are considered, the interaction between locations and timezones may enable interesting approaches.
In Zhang \etal's work~\cite{zhang2015geocoding}, they attempt to disambiguate location mentions with time stamps.
They observe that tweet traffic is fairly low between 2am-5am on weekdays.
When there are several candidate locations (\eg ``Olympia''), they carefully choose one to avoid timezones that place the time stamp in the low traffic window.

\subsection{Summaries and Discussions}
\label{ssec:summary5}

In this section, we review literatures on mentioned location prediction as summarized in Table~\ref{tbl:locationMention}.
Like tweet locations, mentioned locations also depend heavily on tweet content, and slightly on Twitter network and tweet context.
However, we note that mentioned locations does not necessarily imply tweet locations (\eg ``going to Tokyo tomorrow'').
In~\cite{antoine2015portraying}, Antoine \etal use a large volume of tweets to analyze the differences between mentioned locations and tweet locations.
Moreover, due to the definitions, their ground truths are collected differently.
Ground truths for tweet location prediction are obtained by referring to geo-tags of tweets.
Mentioned locations, however, are mostly identified though human annotation~\cite{finin2010annotating}.

Like predicting home and tweet locations, mentioned location prediction also suffers from the noisy and short nature of tweets.
When adopting recognition and disambiguation approaches for formal documents, it is common to involve tweet- and location-specific techniques/information.

Finally, there are a few experimental analysis on tweet NER that are worth noting.
Gelernter \etal~\cite{gelernter2011geo} perform an error analysis on \textit{StanfordNER} for recognizing locations in tweets.
They do not retrain \textit{StanfordNER} with labeled tweets, but use the off-the-shelf version.
Lingad \etal~\cite{lingad2013location} compare a few NER tools on disaster related Twitter data, \eg \textit{StanfordNER}, \textit{OpenNLP}\footnote{http://opennlp.apache.org}, \textit{Yahoo! PlaceMaker}\footnote{http://developer.yahoo.com/geo/placemaker/}, and \textit{TwitterNLP}~\cite{ritter2011named}.
They find that retrained \textit{StanfordNER} outperforms the other competitors.
Liu \etal~\cite{liu2014automatic} also make a similar comparison between \textit{LER} proposed by themselves and other tools.
Besides \textit{StanfordNER} and \textit{TwitterNLP}, they also include \textit{GeoLocator}~\cite{gelernter2013algorithm}, and \textit{UnlockText}\footnote{http://edina.ac.uk/unlock/texts/}.
Derczynski \etal~\cite{derczynski2015analysis} compare tweet NER performances of several systems, but they do not restrict to location entities.

\section{Other Related Problems}
\label{sec:other}

In this section, we review two other problems related to location prediction on Twitter, namely \emph{semantic location prediction} and \emph{point-of-interest recommendation}. We will also try to highlight their differences in terms of definitions, ground truths, and solutions.

\subsection{Semantic Location Prediction}
\label{ssec:semantic}

In Section~\ref{sec:tweetLocation}, we show that many studies depend tweet locations heavily on tweet content.
The underlying assumption is that, if a tweet semantically talks about a location, it is likely to be posted at the venue.
However, people could talk about New York where they visited before but currently locate in Japan.
Thus, semantic locations and tweet locations may not always coincide.
Therefore, some studies focus on predicting semantic locations instead of tweet locations.

Dalvi \etal~\cite{DBLP:conf/wsdm/DalviKP12} investigate matching users' tweets to restaurants in \textit{Yahoo! Local}\footnote{\url{http://local.yahoo.com}, though it is offline now.}.
Those tweets may talk about dishes, service, or ambience of certain restaurants.
They assume that each user has a latent location, and that they are likely to talk about nearby restaurants.
When talking about restaurants, users follow a restaurant-specific bigram language model.
To evaluate their model, they manually annotate hundreds of tweets, where candidate restaurants are suggested by a base system in their previous work.
Zhao \etal~\cite{zhao2016annotating} study matching tweets to general POIs on \textit{Foursquare}.
Different from \cite{DBLP:conf/wsdm/DalviKP12} and other studies on tweet location prediction, they assume that geo-tags of tweets are known and given as input.
Nearby locations with compatible keywords are preferred in the matching.
By introducing dummy locations, their model is capable of identifying the ``no semantic location'' cases.
Evaluations are conducted with thousands of manually annotated tweets.

To sum up, this line of work is characterized by the need of manually annotated ground truth due to the subjective definition of semantic location.
We note that manual annotations take much more efforts to obtain than geo-tags.
Dalvi \etal~\cite{DBLP:conf/wsdm/DalviKP12} and Zhao \etal~\cite{zhao2016annotating} only involve hundreds or thousands of annotated tweets for evaluation respectively.
This could explain why this problem attracts less attention than the three major tasks introduced above.

\subsection{Point-of-Interest Recommendation}
\label{ssec:POIRecommend}

Due to its content-centric nature, Twitter is regarded by users as an ideal platform to share events, emotions, and opinions.
Meanwhile, location-based social networks (LBSNs) like Foursquare, Gowalla, Brightkite, and Yelp concentrate more on POI-centric information.
Besides establishing online friendships, they encourage users to check in, rate, and comment on POIs, as well as keep their information up to date.
The popularity of LBSNs has given rise to abundant studies on POI recommendation.

Due to its popularity~\cite{cheng2011exploring}, Foursquare is adopted by many studies~\cite{ye2010location,gao2012exploring,gao2012gscorr,yang2013sentiment,gao2015content,liu2013point,liu2013learning,noulas2012mining,cheng2013you,DBLP:conf/cikm/YaoZHB17} as data source.
However, Foursquare APIs do not allow access to users' check-in history as reported in many studies.
Luckily, when checking in on Foursquare, users may optionally allow Foursquare to send tweets like ``I'm at [POI] [Foursquare URL of POI].''
By monitoring Twitter streams, researchers manage to accumulate sufficient check-in data for POI recommendation.
This might be the most significant connection between this line of study and Twitter.
In the following, we clarify the differences between POI recommendation and main tasks in this survey.

Judging from the names, POI recommendation focuses on locations at fine-grained POI level.
Moreover, it aims at suggesting POIs that users have never been to, instead of locations that they have connection with~\cite{gao2014personalized}.
A user does not need to write a tweet to get suggested places to visit.
Recommendations are made based on the user's and others' historical data, including check-ins, ratings, and comments, as well as context like the current time and user location.
Finally, evaluation methods are also different: for each user in test set, visited POIs after some checkpoint time or selected samples are masked, predicted, and evaluated.

In terms of solutions, POI recommendations are generally based on collaborative filtering framework.
Although user friendship, content and context are also exploited, they mostly come from LBSNs rather than Twitter.
For friendship, Ye \etal and Gao \etal~\cite{ye2010location,gao2012exploring,gao2012gscorr} employ Foursquare friendship network, while Ying \etal and Cho \etal~\cite{ying2012urban,cho2011friendship} rely on Gowalla and Brightkite networks.
Yang \etal~\cite{yang2013sentiment} claim that Foursquare friendship is not public\footnote{By the time we finish this survey, authorizations from Foursquare users are needed to access their friends via API. However, one can view any user's friend list via a browser.}, and turn to Twitter network.
As for content, check-in tweets do not provide as much textual information as ordinary tweets.
However, several Foursquare-based studies manage to explore user comments~\cite{yang2013sentiment,gao2015content} and POI tags/descriptions~\cite{liu2013point,liu2013learning} in recommendation.
Hasan \etal~\cite{DBLP:conf/kdd/HasanZU13} find that the time of visiting different places depends on types of activities. Such spatio-temporal context is also involved in many other investigations~\cite{noulas2012mining,cheng2013you,gao2012exploring}.

This section is only aimed at clarifying connections and differences between LBSN-based POI recommendation and Twitter-based location prediction.
Due to the scope of this survey, we only involve a small portion of recommendation studies.
Readers may refer to~\cite{gao2014personalized,bao2015recommendations,zhao2016survey} for extensive surveys and~\cite{Liu:2017:EEP} for an experimental evaluation.

\section{Conclusion and Future Work}
\label{sec:Conclusion}

In this survey, we review and summarize techniques of three geolocation problems on Twitter: home location, tweet location, and mentioned location.
Compared with similar problems on formal documents, \ie document geolocation and named entity recognition \& disambiguation, geolocation problems on Twitter face unique challenges and opportunities.
The challenges generally arise from the noisy and short nature of tweet content.
The opportunities, on the other hand, are enabled by the massive Twitter network and rich tweet context.

All the three prediction problems rely heavily on tweet content.
For home and tweet location prediction, techniques could be categorized to the following two classes:
\begin{itemize}
  \item Word-centric methods. They are characterized by identifying local words and modeling spatial word usage.
  Statistical, information theory and heuristic rule-based methods are designed to select location indicative words without supervision.
  Researchers also consider supervised ways to identify local words based on manual features and annotations.
  When modeling spatial word usage, direct estimations from data may suffer from sparsity problem.
  Therefore, multiple smoothing techniques are proposed.
  \item Location-centric methods. They are characterized by constructing pseudo-documents or classifiers for locations.
  Pseudo-documents construction are essential for information-retrieval-inspired approaches.
  Similar to spatial word usage, language models for pseudo-documents also require smoothing.
  However, geographical smoothing techniques, \eg Gaussian model and grid-based smoothing, are not applicable.
  For tweet location prediction, classification methods are rarely adopted because it is usually at fine-grained POI level.
\end{itemize}
As for mentioned location, efforts on recognition address the noisy-content challenge by sophisticated features and comprehensive gazetteers.
Collective disambiguation is employed to relieve the information scarcity brought by short tweets.
Jointly optimizing both recognition and disambiguation components is also advocated in some studies.

As a significant feature of the platform, Twitter network plays a key role in home location prediction.
Various hypotheses have been made on the connections between friendship and home proximity. Inspired by Backstrom \etal~\cite{backstrom2010find}, many works try to formulate the relationship between the probability of friendship and home location distance. However, the indication is not very strong on Twitter.
To fix this issue, social-closeness-based methods are proposed to differentiate noisy friendship.
Explicit factors like friends with interactions are employed as useful information to predict home proximity.
Implicit factors like influence scope are captured by sophisticated models.
Finally, we note that Twitter network causes the predictions for different users to depend on each other.
Therefore, it is necessary to involve global inference approaches.

Though short in length, tweets are accompanied with rich context.
Those include timestamps and geo-tags associated with tweets, as well as various attributes in user profiles.
Among them, temporal information like tweet timestamps and user-declared timezones are effective in implying tweet and home locations at coarse-grained granularity.
Geo-tags and timestamps are also proven informative for disambiguating mentioned locations and other types of entities.
Finally, we relate semantic location prediction for tweets and LBSN-based POI recommendation.
We note that spatio-temporal factors are modeled in a more sophisticated manner in LBSN-based POI recommendation.

Geolocation is not only tackled on Twitter, but also many other platforms like Facebook~\cite{backstrom2010find}, Foursquare~\cite{jurgens2013s}, Gowalla~\cite{kong2014spot}, \etc The prediction models proposed based on Twitter can also be adapted to other social media sites, while might require some changes. But before considering model adaptations, we need to be clear on whether the three geolocation problems on Twitter, \ie prediction of home location, tweet location and mentioned location, are applicable to the target platform or not. For example, tweet and mentioned location prediction on some image and video sharing platforms like Instagram and Pinterest may not be applicable. Next, the differences of available information, \ie content, network, context, between the target platform and Twitter is another main consideration adapting the models on Twitter to other platforms. An example is that the friendship relationship on Facebook is bidirectional, but is unidirectional on Twitter.

At last, we would like to suggest some future directions. 	First, deep learning methods demonstrate great ability of learning feature representations automatically. A few recent works~\cite{DBLP:conf/acl/RahimiCB17,DBLP:conf/acl/MiuraTTO17}  tried to apply neural network models directly to geolocation problems on Twitter and achieved some progress. Appropriate combination of Twitter properties and neural networks on geolocation deserves further research.

Second, most of current reviewed methods mainly focus on content information. The usage of network and context is not well investigated, especially for tweet and mentioned location prediction. In addition, the interactions among content, context, and network are not well analyzed. Most of current methods assume them to be independent features and combine them in a linear fashion. Joint modeling and exploiting of those factors could be a possible direction.

Third, data sparsity is a major issue for geolocation problem, especially for tweet and mentioned location prediction. Effective methods to augment useful information  leave a big room to improve. Reliable images or cross-platform information might help to improve the performance. The exploration of appropriate approaches to leverage auxiliary knowledge also need more research.

\section*{Acknowledgments}
Xin Zheng is in the SAP Industrial Ph.D Program, partially funded by the Economic Development Board and the National Research Foundation of Singapore.  This work was partially supported by Singapore Ministry of Education Academic Research Fund MOE2014-T2-2-066.

\ifCLASSOPTIONcaptionsoff
  \newpage
\fi


\begin{thebibliography}{100}
\providecommand{\url}[1]{#1}
\csname url@samestyle\endcsname
\providecommand{\newblock}{\relax}
\providecommand{\bibinfo}[2]{#2}
\providecommand{\BIBentrySTDinterwordspacing}{\spaceskip=0pt\relax}
\providecommand{\BIBentryALTinterwordstretchfactor}{4}
\providecommand{\BIBentryALTinterwordspacing}{\spaceskip=\fontdimen2\font plus
\BIBentryALTinterwordstretchfactor\fontdimen3\font minus
  \fontdimen4\font\relax}
\providecommand{\BIBforeignlanguage}[2]{{%
\expandafter\ifx\csname l@#1\endcsname\relax
\typeout{** WARNING: IEEEtran.bst: No hyphenation pattern has been}%
\typeout{** loaded for the language `#1'. Using the pattern for}%
\typeout{** the default language instead.}%
\else
\language=\csname l@#1\endcsname
\fi
#2}}
\providecommand{\BIBdecl}{\relax}
\BIBdecl

\bibitem{cheng2010you}
Z.~Cheng, J.~Caverlee, and K.~Lee, ``You are where you tweet: a content-based
  approach to geo-locating twitter users,'' in \emph{Proc. {ACM} Conf. on
  Information and Knowledge Management}, 2010, pp. 759--768.

\bibitem{yuan2013and}
Q.~Yuan, G.~Cong, Z.~Ma, A.~Sun, and N.~M. Thalmann, ``Who, where, when and
  what: discover spatio-temporal topics for twitter users,'' in \emph{Proc.
  {ACM} Conf. on Knowledge Discovery and Data Mining}, 2013, pp. 605--613.

\bibitem{noulas2012mining}
A.~Noulas, S.~Scellato, N.~Lathia, and C.~Mascolo, ``Mining user mobility
  features for next place prediction in location-based services,'' in
  \emph{Proc. {IEEE} Int. Conf. on Data Mining}, 2012, pp. 1038--1043.

\bibitem{rakesh2013location}
V.~Rakesh, C.~K. Reddy, D.~Singh, and M.~Ramachandran, ``Location-specific
  tweet detection and topic summarization in twitter,'' in \emph{Proc. Advances
  in Social Networks Analysis and Mining}, 2013, pp. 1441--1444.

\bibitem{ao2014estimating}
J.~Ao, P.~Zhang, and Y.~Cao, ``Estimating the locations of emergency events
  from twitter streams,'' in \emph{Proc. Int. Conf. on Information Technology
  and Quantitative Management}, 2014, pp. 731--739.

\bibitem{lingad2013location}
J.~Lingad, S.~Karimi, and J.~Yin, ``Location extraction from disaster-related
  microblogs,'' in \emph{Proc. World Wide Web Conf. Companion Volume}, 2013,
  pp. 1017--1020.

\bibitem{cheng2013content}
Z.~Cheng, J.~Caverlee, and K.~Lee, ``A content-driven framework for geolocating
  microblog users,'' \emph{ACM Trans. on Intelligent Systems and Technology},
  vol.~4, no.~1, p.~2, 2013.

\bibitem{hecht2011tweets}
B.~Hecht, L.~Hong, B.~Suh, and E.~H. Chi, ``Tweets from justin bieber's heart:
  the dynamics of the location field in user profiles,'' in \emph{Proc. Conf.
  on Human Factors in Computing Systems}, 2011, pp. 237--246.

\bibitem{ryoo2014inferring}
K.~Ryoo and S.~Moon, ``Inferring twitter user locations with 10 km accuracy,''
  in \emph{Proc. World Wide Web Conf. Companion Volume}, 2014, pp. 643--648.

\bibitem{hawelka2014geo}
B.~Hawelka, I.~Sitko, E.~Beinat, S.~Sobolevsky, P.~Kazakopoulos, and C.~Ratti,
  ``Geo-located twitter as proxy for global mobility patterns,''
  \emph{Cartography and Geographic Information Science}, vol.~41, no.~3, pp.
  260--271, 2014.

\bibitem{priedhorsky2014inferring}
R.~Priedhorsky, A.~Culotta, and S.~Y. Del~Valle, ``Inferring the origin
  locations of tweets with quantitative confidence,'' in \emph{Proc. ACM Conf.
  on Computer Supported Cooperative Work and Social Computing}, 2014, pp.
  1523--1536.

\bibitem{Wing:2011:SSD:2002472.2002593}
B.~P. Wing and J.~Baldridge, ``Simple supervised document geolocation with
  geodesic grids,'' in \emph{Proc. Annual Meeting of the Association for
  Computational Linguistics: Human Language Technologies - Volume 1}, 2011, pp.
  955--964.

\bibitem{wing2014hierarchical}
B.~Wing and J.~Baldridge, ``Hierarchical discriminative classification for
  text-based geolocation,'' in \emph{Proc. Conf. on Empirical Methods in
  Natural Language Processing}, 2014, pp. 336--348.

\bibitem{roller2012supervised}
S.~Roller, M.~Speriosu, S.~Rallapalli, B.~Wing, and J.~Baldridge, ``Supervised
  text-based geolocation using language models on an adaptive grid,'' in
  \emph{Proc. Joint Conf. on Empirical Methods in Natural Language Processing
  and Computational Natural Language Learning}, 2012, pp. 1500--1510.

\bibitem{amitay2004web}
E.~Amitay, N.~Har'El, R.~Sivan, and A.~Soffer, ``Web-a-where: geotagging web
  content,'' in \emph{Proc. ACM SIGIR Conf. on Research and Development in
  Information Retrieval}, 2004, pp. 273--280.

\bibitem{Zong2005JCDL}
W.~Zong, D.~Wu, A.~Sun, E.-P. Lim, and D.~H.-L. Goh, ``On assigning place names
  to geography related web pages,'' in \emph{Proc. ACM/IEEE-CS Joint Conf. on
  Digital Libraries}, 2005, pp. 354--362.

\bibitem{woodruff1994gipsy}
A.~Woodruff and C.~Plaunt, ``Gipsy: Automated geographic indexing of text
  documents,'' \emph{J. the American Society for Information Science}, vol.~45,
  no.~9, pp. 645--655, 1994.

\bibitem{nadeau2007survey}
D.~Nadeau and S.~Sekine, ``A survey of named entity recognition and
  classification,'' \emph{Lingvisticae Investigationes}, vol.~30, no.~1, pp.
  3--26, 2007.

\bibitem{DBLP:journals/tkde/ShenWH15}
W.~Shen, J.~Wang, and J.~Han, ``Entity linking with a knowledge base: Issues,
  techniques, and solutions,'' \emph{{IEEE} Trans. Knowl. Data Eng.}, vol.~27,
  no.~2, pp. 443--460, 2015.

\bibitem{imran2014processing}
M.~Imran, C.~Castillo, F.~Diaz, and S.~Vieweg, ``Processing social media
  messages in mass emergency: A survey,'' \emph{ACM Computing Surveys},
  vol.~47, no.~4, p.~67, 2015.

\bibitem{TGIS:TGIS12212}
F.~Melo and B.~Martins, ``Automated geocoding of textual documents: {A} survey
  of current approaches,'' \emph{Trans. {GIS}}, vol.~21, no.~1, pp. 3--38,
  2017.

\bibitem{DBLP:journals/jis/AjaoHL15}
O.~Ajao, J.~Hong, and W.~Liu, ``A survey of location inference techniques on
  twitter,'' \emph{J. Information Science}, vol.~41, no.~6, pp. 855--864, 2015.

\bibitem{mcgee2011geographic}
J.~McGee, J.~A. Caverlee, and Z.~Cheng, ``A geographic study of tie strength in
  social media,'' in \emph{Proc. {ACM} Conf. on Information and Knowledge
  Management}, 2011, pp. 2333--2336.

\bibitem{mcgee2013location}
J.~McGee, J.~Caverlee, and Z.~Cheng, ``Location prediction in social media
  based on tie strength,'' in \emph{Proc. {ACM} Conf. on Information and
  Knowledge Management}, 2013, pp. 459--468.

\bibitem{compton2014geotagging}
R.~Compton, D.~Jurgens, and D.~Allen, ``Geotagging one hundred million twitter
  accounts with total variation minimization,'' in \emph{Proc. {IEEE} Int.
  Conf. on Big Data}, 2014, pp. 393--401.

\bibitem{jurgens2013s}
D.~Jurgens, ``That's what friends are for: Inferring location in online social
  media platforms based on social relationships,'' in \emph{Proc. Int. Conf. on
  Weblogs and Social Media}, 2013.

\bibitem{yamaguchi2014online}
Y.~Yamaguchi, T.~Amagasa, H.~Kitagawa, and Y.~Ikawa, ``Online user location
  inference exploiting spatiotemporal correlations in social streams,'' in
  \emph{Proc. {ACM} Conf. on Information and Knowledge Management}, 2014, pp.
  1139--1148.

\bibitem{DBLP:conf/coling/HanCB12}
B.~Han, P.~Cook, and T.~Baldwin, ``Geolocation prediction in social media data
  by finding location indicative words,'' in \emph{Proc. Conf. on Computational
  Linguistics: Technical Papers}, 2012, pp. 1045--1062.

\bibitem{schulz2013multi}
A.~Schulz, A.~Hadjakos, H.~Paulheim, J.~Nachtwey, and M.~M{\"u}hlh{\"a}user,
  ``A multi-indicator approach for geolocalization of tweets,'' in \emph{Proc.
  Int. Conf. on Weblogs and Social Media}, 2013.

\bibitem{li2012towards}
R.~Li, S.~Wang, H.~Deng, R.~Wang, and K.~C.-C. Chang, ``Towards social user
  profiling: unified and discriminative influence model for inferring home
  locations,'' in \emph{Proc. {ACM} Conf. on Knowledge Discovery and Data
  Mining}, 2012, pp. 1023--1031.

\bibitem{han2013stacking}
B.~Han, P.~Cook, and T.~Baldwin, ``A stacking-based approach to twitter user
  geolocation prediction,'' in \emph{Proc. Annual Meeting of the Association
  for Computational Linguistics System Demonstrations}, 2013, pp. 7--12.

\bibitem{flatow2015accuracy}
D.~Flatow, M.~Naaman, K.~E. Xie, Y.~Volkovich, and Y.~Kanza, ``On the accuracy
  of hyper-local geotagging of social media content,'' in \emph{Proc. {ACM}
  Conf. on Web Search and Data Mining}, 2015, pp. 127--136.

\bibitem{DBLP:journals/tkde/LaereQSD14}
O.~V. Laere, J.~A. Quinn, S.~Schockaert, and B.~Dhoedt, ``Spatially aware term
  selection for geotagging,'' \emph{{IEEE} Trans. Knowl. Data Eng.}, vol.~26,
  no.~1, pp. 221--234, 2014.

\bibitem{silverman1986density}
B.~W. Silverman, \emph{Density estimation for statistics and data
  analysis}.\hskip 1em plus 0.5em minus 0.4em\relax CRC press, 1986, vol.~26.

\bibitem{ripley2005spatial}
B.~D. Ripley, \emph{Spatial statistics}.\hskip 1em plus 0.5em minus 0.4em\relax
  John Wiley \& Sons, 2005, vol. 575.

\bibitem{ren2012you}
K.~Ren, S.~Zhang, and H.~Lin, ``Where are you settling down: Geo-locating
  twitter users based on tweets and social networks,'' in \emph{Proc. Asia
  Information Retrieval Symposium}, 2012, pp. 150--161.

\bibitem{mahmud2012tweet}
J.~Mahmud, J.~Nichols, and C.~Drews, ``Where is this tweet from? inferring home
  locations of twitter users,'' in \emph{Proc. Int. Conf. on Weblogs and Social
  Media}, 2012.

\bibitem{DBLP:journals/jair/HanCB14}
B.~Han, P.~Cook, and T.~Baldwin, ``Text-based twitter user geolocation
  prediction,'' \emph{J. Artif. Intell. Res.}, vol.~49, pp. 451--500, 2014.

\bibitem{backstrom2008spatial}
L.~Backstrom, J.~Kleinberg, R.~Kumar, and J.~Novak, ``Spatial variation in
  search engine queries,'' in \emph{Proc. Conf. World Wide Web}, 2008.

\bibitem{li2012multiple}
R.~Li, S.~Wang, and K.~C.-C. Chang, ``Multiple location profiling for users and
  relationships from social network and content,'' \emph{{PVLDB}}, vol.~5,
  no.~11, pp. 1603--1614, 2012.

\bibitem{chang2012phillies}
H.~wen Chang, D.~Lee, M.~Eltaher, and J.~Lee, ``@ phillies tweeting from
  philly? predicting twitter user locations with spatial word usage,'' in
  \emph{Proc. Conf. on Advances in Social Networks Analysis and Mining}, 2012,
  pp. 111--118.

\bibitem{DBLP:conf/naacl/RahimiVCB15}
A.~Rahimi, D.~Vu, T.~Cohn, and T.~Baldwin, ``Exploiting text and network
  context for geolocation of social media users,'' in \emph{Proc. Conf. of the
  North American Chapter of the Association for Computational Linguistics:
  Human Language Technologies}, 2015, pp. 1362--1367.

\bibitem{tibshirani1996regression}
R.~Tibshirani, ``Regression shrinkage and selection via the lasso,'' \emph{J.
  the Royal Statistical Society. Series B (Methodological)}, pp. 267--288,
  1996.

\bibitem{DBLP:conf/icwsm/ChaGK15}
M.~Cha, Y.~Gwon, and H.~T. Kung, ``Twitter geolocation and regional
  classification via sparse coding,'' in \emph{Proc. Int. Conf. on Web and
  Social Media}, 2015, pp. 582--585.

\bibitem{mahmud2014home}
J.~Mahmud, J.~Nichols, and C.~Drews, ``Home location identification of twitter
  users,'' \emph{ACM Trans. on Intelligent Systems and Technology}, vol.~5,
  no.~3, pp. 47:1--47:21, 2014.

\bibitem{silla2011survey}
C.~N. Silla~Jr and A.~A. Freitas, ``A survey of hierarchical classification
  across different application domains,'' \emph{Data Mining and Knowledge
  Discovery}, vol.~22, no. 1-2, pp. 31--72, 2011.

\bibitem{ponte1998language}
J.~M. Ponte and W.~B. Croft, ``A language modeling approach to information
  retrieval,'' in \emph{Proc. {ACM} {SIGIR} Conf. on Research and Development
  in Information Retrieval}, 1998, pp. 275--281.

\bibitem{good1953population}
I.~J. Good, ``The population frequencies of species and the estimation of
  population parameters,'' \emph{Biometrika}, vol.~40, no. 3-4, pp. 237--264,
  1953.

\bibitem{DBLP:conf/aclnut/MiuraTTO16}
Y.~Miura, M.~Taniguchi, T.~Taniguchi, and T.~Ohkuma, ``A simple scalable neural
  networks based model for geolocation prediction in twitter,'' in \emph{Proc.
  Workshop on Noisy User-generated Text}, 2016, pp. 235--239.

\bibitem{DBLP:conf/acl/MiuraTTO17}
------, ``Unifying text, metadata, and user network representations with a
  neural network for geolocation prediction,'' in \emph{Proc. Annual Meeting of
  the Association for Computational Linguistics}, 2017, pp. 1260--1272.

\bibitem{DBLP:conf/acl/RahimiCB17}
A.~Rahimi, T.~Cohn, and T.~Baldwin, ``A neural model for user geolocation and
  lexical dialectology,'' in \emph{Proc. Annual Meeting of the Association for
  Computational Linguistics, Volume 2: Short Papers}, 2017, pp. 209--216.

\bibitem{mcpherson2001birds}
M.~McPherson, L.~Smith-Lovin, and J.~M. Cook, ``Birds of a feather: Homophily
  in social networks,'' \emph{Annual review of sociology}, vol.~27, no.~1, pp.
  415--444, 2001.

\bibitem{davis2011inferring}
C.~A. Davis~Jr, G.~L. Pappa, D.~R.~R. de~Oliveira, and F.~de~L~Arcanjo,
  ``Inferring the location of twitter messages based on user relationships,''
  \emph{Trans. GIS}, vol.~15, no.~6, pp. 735--751, 2011.

\bibitem{DBLP:journals/ijon/RodriguesAPOM16}
E.~C. Rodrigues, R.~Assun{\c{c}}{\~{a}}o, G.~L. Pappa, D.~R.~R. Oliveira, and
  W.~M. Jr., ``Exploring multiple evidence to infer users' location in
  twitter,'' \emph{Neurocomputing}, vol. 171, pp. 30--38, 2016.

\bibitem{DBLP:series/acvpr/978-1-84800-278-4}
S.~Z. Li, \emph{Markov Random Field Modeling in Image Analysis}, ser. Advances
  in Pattern Recognition.\hskip 1em plus 0.5em minus 0.4em\relax Springer,
  2009.

\bibitem{backstrom2010find}
L.~Backstrom, E.~Sun, and C.~Marlow, ``Find me if you can: improving
  geographical prediction with social and spatial proximity,'' in \emph{Proc.
  Conf. World Wide Web}, 2010.

\bibitem{kong2014spot}
L.~Kong, Z.~Liu, and Y.~Huang, ``{SPOT:} locating social media users based on
  social network context,'' \emph{{PVLDB}}, vol.~7, no.~13, pp. 1681--1684,
  2014.

\bibitem{kossinets2006empirical}
G.~Kossinets and D.~J. Watts, ``Empirical analysis of an evolving social
  network,'' \emph{Science}, vol. 311, no. 5757, pp. 88--90, 2006.

\bibitem{rout2013s}
D.~Rout, K.~Bontcheva, D.~Preo{\c{t}}iuc-Pietro, and T.~Cohn, ``Where's
  @wally?: a classification approach to geolocating users based on their social
  ties,'' in \emph{Proc. {ACM} Conf. on Hypertext and Social Media}, 2013, pp.
  11--20.

\bibitem{kwak2010twitter}
H.~Kwak, C.~Lee, H.~Park, and S.~Moon, ``What is twitter, a social network or a
  news media?'' in \emph{Proc. Conf. World Wide Web}, 2010, pp. 591--600.

\bibitem{eisenstein2010latent}
J.~Eisenstein, B.~O'Connor, N.~A. Smith, and E.~P. Xing, ``A latent variable
  model for geographic lexical variation,'' in \emph{Proc. Conf. on Empirical
  Methods in Natural Language Processing}, 2010, pp. 1277--1287.

\bibitem{yamaguchi2013landmark}
Y.~Yamaguchi, T.~Amagasa, and H.~Kitagawa, ``Landmark-based user location
  inference in social media,'' in \emph{Proc. Conf. on Online Social Networks},
  2013, pp. 223--234.

\bibitem{DBLP:conf/acl/RahimiCB15}
A.~Rahimi, T.~Cohn, and T.~Baldwin, ``Twitter user geolocation using a unified
  text and network prediction model,'' in \emph{Proc. Meeting of the
  Association for Computational Linguistics and the Joint Conf. on Natural
  Language Processing of the Asian Federation of Natural Language Processing},
  2015, pp. 630--636.

\bibitem{eisenstein2011sparse}
J.~Eisenstein, A.~Ahmed, and E.~P. Xing, ``Sparse additive generative models of
  text,'' in \emph{Proc. Int. Conf. on Machine Learning}, 2011, pp. 1041--1048.

\bibitem{DBLP:conf/asunam/EfstathiadesAPD15}
H.~Efstathiades, D.~Antoniades, G.~Pallis, and M.~D. Dikaiakos,
  ``Identification of key locations based on online social network activity,''
  in \emph{Proc. {IEEE/ACM} Conf. on Advances in Social Networks Analysis and
  Mining}, 2015, pp. 218--225.

\bibitem{DBLP:conf/ht/PoulstonSB17}
A.~Poulston, M.~Stevenson, and K.~Bontcheva, ``Hyperlocal home location
  identification of twitter profiles,'' in \emph{Proc. {ACM} Conf. on Hypertext
  and Social Media}, 2017, pp. 45--54.

\bibitem{DBLP:conf/emnlp/RahimiBC17}
A.~Rahimi, T.~Baldwin, and T.~Cohn, ``Continuous representation of location for
  geolocation and lexical dialectology using mixture density networks,'' in
  \emph{Proc. Conf. on Empirical Methods in Natural Language Processing}, 2017,
  pp. 167--176.

\bibitem{scellato2011socio}
S.~Scellato, A.~Noulas, R.~Lambiotte, and C.~Mascolo, ``Socio-spatial
  properties of online location-based social networks,'' in \emph{Proc. Int.
  Conf. on Weblogs and Social Media}, 2011.

\bibitem{volkovich2012length}
Y.~Volkovich, S.~Scellato, D.~Laniado, C.~Mascolo, and A.~Kaltenbrunner, ``The
  length of bridge ties: Structural and geographic properties of online social
  interactions,'' in \emph{Proc. Int. Conf. on Weblogs and Social Media}, 2012.

\bibitem{zhu2002learning}
X.~Zhu and Z.~Ghahramani, ``Learning from labeled and unlabeled data with label
  propagation,'' Citeseer, Tech. Rep., 2002.

\bibitem{richtarik2016parallel}
P.~Richt{\'a}rik and M.~Tak{\'a}{\v{c}}, ``Parallel coordinate descent methods
  for big data optimization,'' \emph{Mathematical Programming}, vol. 156, no.
  1-2, pp. 433--484, 2016.

\bibitem{DBLP:journals/ml/AndrieuFDJ03}
C.~Andrieu, N.~de~Freitas, A.~Doucet, and M.~I. Jordan, ``An introduction to
  {MCMC} for machine learning,'' \emph{Machine Learning}, vol.~50, no. 1-2, pp.
  5--43, 2003.

\bibitem{DBLP:conf/sigir/ChengCBB14}
Z.~Cheng, J.~Caverlee, H.~Barthwal, and V.~Bachani, ``Who is the barbecue king
  of texas?: a geo-spatial approach to finding local experts on twitter,'' in
  \emph{Proc. {ACM} {SIGIR} Conf. on Research and Development in Information
  Retrieval}, 2014, pp. 335--344.

\bibitem{jurgens2015geolocation}
D.~Jurgens, T.~Finethy, J.~McCorriston, Y.~T. Xu, and D.~Ruths, ``Geolocation
  prediction in twitter using social networks: A critical analysis and review
  of current practice,'' in \emph{Proc. Int. Conf. on Web and Social Media},
  2015, pp. 188--197.

\bibitem{java2009we}
A.~Java, X.~Song, T.~Finin, and B.~Tseng, ``Why we twitter: An analysis of a
  microblogging community,'' in \emph{Proc. Workshop on Knowledge Discovery on
  the Web and Workshop on Social Networks Analysis}, 2007, pp. 118--138.

\bibitem{graham2014world}
M.~Graham, S.~A. Hale, and D.~Gaffney, ``Where in the world are you?
  geolocation and language identification in twitter,'' \emph{The Professional
  Geographer}, vol.~66, no.~4, pp. 568--578, 2014.

\bibitem{DBLP:conf/icwsm/ChongL17}
W.~Chong and E.~Lim, ``Exploiting contextual information for fine-grained tweet
  geolocation,'' in \emph{Proc. Int. Conf. on Web and Social Media}, 2017, pp.
  488--491.

\bibitem{DBLP:conf/cikm/ChongL17}
------, ``Tweet geolocation: Leveraging location, user and peer signals,'' in
  \emph{Proc. ACM Conf. on Information and Knowledge Management}, 2017, pp.
  1279--1288.

\bibitem{kinsella2011m}
S.~Kinsella, V.~Murdock, and N.~O'Hare, ``I'm eating a sandwich in glasgow:
  modeling locations with tweets,'' in \emph{Proc. {CIKM} Workshop on Search
  and Mining User-Generated Contents}, 2011, pp. 61--68.

\bibitem{Zhai:2001:SSM:383952.384019}
C.~Zhai and J.~D. Lafferty, ``A study of smoothing methods for language models
  applied to ad hoc information retrieval,'' in \emph{Proc. {ACM} {SIGIR} Conf.
  on Research and Development in Information Retrieval}, 2001, pp. 334--342.

\bibitem{Li2011the}
W.~Li, P.~Serdyukov, A.~P. de~Vries, C.~Eickhoff, and M.~Larson, ``The where in
  the tweet,'' in \emph{Proc. {ACM} Conf. on Information and Knowledge
  Management}, 2011, pp. 2473--2476.

\bibitem{DBLP:conf/mobiquitous/LeeGSL14}
K.~Lee, R.~K. Ganti, M.~Srivatsa, and L.~Liu, ``When twitter meets foursquare:
  tweet location prediction using foursquare,'' in \emph{Proc. Conf. on Mobile
  and Ubiquitous Systems: Computing, Networking and Services}, 2014, pp.
  198--207.

\bibitem{DBLP:conf/acl/ChenG96}
S.~F. Chen and J.~Goodman, ``An empirical study of smoothing techniques for
  language modeling,'' in \emph{Proc. Meeting of the Association for
  Computational Linguistics}, 1996, pp. 310--318.

\bibitem{DBLP:conf/cikm/Liu016}
Z.~Liu and Y.~Huang, ``Where are you tweeting?: {A} context and user movement
  based approach,'' in \emph{Proc. {ACM} Conf. on Information and Knowledge
  Management}, 2016, pp. 1949--1952.

\bibitem{DBLP:conf/aaai/HuldenSF15}
M.~Hulden, M.~Silfverberg, and J.~Francom, ``Kernel density estimation for
  text-based geolocation,'' in \emph{Proc. {AAAI} Conf. on Artificial
  Intelligence}, 2015, pp. 145--150.

\bibitem{DBLP:conf/naacl/DredzeOK16}
M.~Dredze, M.~Osborne, and P.~Kambadur, ``Geolocation for twitter: Timing
  matters,'' in \emph{Proc. Conf. of the North American Chapter of the
  Association for Computational Linguistics: Human Language Technologies},
  2016, pp. 1064--1069.

\bibitem{DBLP:conf/bigdataconf/CaoCJY15}
B.~Cao, F.~Chen, D.~Joshi, and P.~S. Yu, ``Inferring crowd-sourced venues for
  tweets,'' in \emph{2015 {IEEE} Int. Conf. on Big Data}, 2015, pp. 639--648.

\bibitem{hahmann2014twitter}
S.~Hahmann, R.~S. Purves, and D.~Burghardt, ``Twitter location (sometimes)
  matters: Exploring the relationship between georeferenced tweet content and
  nearby feature classes,'' \emph{J. Spatial Information Science}, vol. 2014,
  no.~9, pp. 1--36, 2014.

\bibitem{DBLP:journals/corr/IsoWA17}
H.~Iso, S.~Wakamiya, and E.~Aramaki, ``Density estimation for geolocation via
  convolutional mixture density network,'' \emph{CoRR}, vol. abs/1705.02750,
  2017.

\bibitem{hong2012discovering}
L.~Hong, A.~Ahmed, S.~Gurumurthy, A.~J. Smola, and K.~Tsioutsiouliklis,
  ``Discovering geographical topics in the twitter stream,'' in \emph{Proc.
  Conf. World Wide Web}, 2012, pp. 769--778.

\bibitem{chen2013interest}
Y.~Chen, J.~Zhao, X.~Hu, X.~Zhang, Z.~Li, and T.-S. Chua, ``From interest to
  function: Location estimation in social media,'' in \emph{Proc. {AAAI} Conf.
  on Artificial Intelligence}, 2013.

\bibitem{sadilek2012finding}
A.~Sadilek, H.~Kautz, and J.~P. Bigham, ``Finding your friends and following
  them to where you are,'' in \emph{Proc. Conf. on Web Search and Data Mining},
  2012, pp. 723--732.

\bibitem{DBLP:conf/icaart/GalalE16}
A.~Galal and A.~El{-}Korany, ``Enabling semantic user context to enhance
  twitter location prediction,'' in \emph{Proc. Int. Conf. on Agents and
  Artificial Intelligence, Volume 1}, 2016, pp. 223--230.

\bibitem{wang2007mining}
C.~Wang, J.~Wang, X.~Xie, and W.-Y. Ma, ``Mining geographic knowledge using
  location aware topic model,'' in \emph{Proc. {ACM} Workshop On Geographic
  Information Retrieval}, 2007, pp. 65--70.

\bibitem{mei2006probabilistic}
Q.~Mei, C.~Liu, H.~Su, and C.~Zhai, ``A probabilistic approach to
  spatiotemporal theme pattern mining on weblogs,'' in \emph{Proc. Conf. on
  World Wide Web}, 2006, pp. 533--542.

\bibitem{maceachren2011senseplace2}
A.~M. MacEachren, A.~Jaiswal, A.~C. Robinson, S.~Pezanowski, A.~Savelyev,
  P.~Mitra, X.~Zhang, and J.~Blanford, ``Senseplace2: Geotwitter analytics
  support for situational awareness,'' in \emph{Proc. {IEEE} Conf. on Visual
  Analytics Science and Technology}, 2011, pp. 181--190.

\bibitem{ritter2011named}
A.~Ritter, S.~Clark, Mausam, and O.~Etzioni, ``Named entity recognition in
  tweets: An experimental study,'' in \emph{Proc. Conf. on Empirical Methods in
  Natural Language Processing}, 2011, pp. 1524--1534.

\bibitem{liu2011recognizing}
X.~Liu, S.~Zhang, F.~Wei, and M.~Zhou, ``Recognizing named entities in
  tweets,'' in \emph{Proc. Annual Meeting of the Association for Computational
  Linguistics}, 2011, pp. 359--367.

\bibitem{liu2013named}
X.~Liu, F.~Wei, S.~Zhang, and M.~Zhou, ``Named entity recognition for tweets,''
  \emph{ACM Trans. on Intelligent Systems and Technology}, vol.~4, no.~1, p.~3,
  2013.

\bibitem{li2012twiner}
C.~Li, J.~Weng, Q.~He, Y.~Yao, A.~Datta, A.~Sun, and B.-S. Lee, ``Twiner: named
  entity recognition in targeted twitter stream,'' in \emph{Proc. {ACM} SIGIR
  Conf. on Research and Development in Information Retrieval}, 2012, pp.
  721--730.

\bibitem{malmasi2015location}
S.~Malmasi and M.~Dras, ``Location mention detection in tweets and
  microblogs,'' in \emph{Proc. Conf. of the Pacific Association for
  Computaitonal Linguistics}, 2015, pp. 123--134.

\bibitem{gelernter2013cross}
J.~Gelernter and W.~Zhang, ``Cross-lingual geo-parsing for non-structured
  data,'' in \emph{Proc. Workshop on Geographic Information Retrieval}, 2013,
  pp. 64--71.

\bibitem{gelernter2013algorithm}
J.~Gelernter and S.~Balaji, ``An algorithm for local geoparsing of microtext,''
  \emph{GeoInformatica}, vol.~17, no.~4, pp. 635--667, 2013.

\bibitem{li2014fine}
C.~Li and A.~Sun, ``Fine-grained location extraction from tweets with temporal
  awareness,'' in \emph{Proc. {ACM} SIGIR Conf. on Research and Development in
  Information Retrieval}, 2014, pp. 43--52.

\bibitem{liextracting}
------, ``Extracting fine-grained location with temporal awareness in tweets: A
  two-stage approach,'' \emph{J. the Association for Information Science and
  Technology}, vol.~68, no.~7, pp. 1652--1670, 2017.

\bibitem{zhang2015geocoding}
W.~Zhang and J.~Gelernter, ``Geocoding location expressions in twitter
  messages: A preference learning method,'' \emph{J. Spatial Information
  Science}, vol.~9, no.~1, pp. 37--70, 2014.

\bibitem{DBLP:conf/www/JiSCH16}
Z.~Ji, A.~Sun, G.~Cong, and J.~Han, ``Joint recognition and linking of
  fine-grained locations from tweets,'' in \emph{Proc. Conf. World Wide Web},
  2016, pp. 1271--1281.

\bibitem{li2014effective}
G.~Li, J.~Hu, J.~Feng, and K.-l. Tan, ``Effective location identification from
  microblogs,'' in \emph{Proc. IEEE Int. Conf. on Data Eng.}\hskip 1em plus
  0.5em minus 0.4em\relax IEEE, 2014, pp. 880--891.

\bibitem{DBLP:conf/kdd/ShenWLW13}
W.~Shen, J.~Wang, P.~Luo, and M.~Wang, ``Linking named entities in tweets with
  knowledge base via user interest modeling,'' in \emph{Proc. {ACM} Conf. on
  Knowledge Discovery and Data Mining}, 2013, pp. 68--76.

\bibitem{DBLP:conf/naacl/GuoCK13}
S.~Guo, M.~Chang, and E.~Kiciman, ``To link or not to link? {A} study on
  end-to-end tweet entity linking,'' in \emph{HLT-NAACL}, 2013, pp. 1020--1030.

\bibitem{DBLP:conf/sigmod/HuaZZ15}
W.~Hua, K.~Zheng, and X.~Zhou, ``Microblog entity linking with social temporal
  context,'' in \emph{Proc. {ACM} {SIGMOD} Conf. on Management of Data}, 2015,
  pp. 1761--1775.

\bibitem{DBLP:journals/tacl/FangC14}
Y.~Fang and M.~Chang, ``Entity linking on microblogs with spatial and temporal
  signals,'' \emph{{TACL}}, vol.~2, pp. 259--272, 2014.

\bibitem{lafferty2001conditional}
J.~Lafferty, A.~McCallum, and F.~C. Pereira, ``Conditional random fields:
  Probabilistic models for segmenting and labeling sequence data,'' in
  \emph{Proc. Int. Conf. on Machine Learning}, 2001, pp. 282--289.

\bibitem{ratinov2009design}
L.~Ratinov and D.~Roth, ``Design challenges and misconceptions in named entity
  recognition,'' in \emph{Proc. Conf. on Computational Natural Language
  Learning}, 2009, pp. 147--155.

\bibitem{brown1992class}
P.~F. Brown, P.~V. Desouza, R.~L. Mercer, V.~J.~D. Pietra, and J.~C. Lai,
  ``Class-based n-gram models of natural language,'' \emph{Computational
  Linguistics}, vol.~18, no.~4, pp. 467--479, 1992.

\bibitem{lieberman2010geotagging}
M.~D. Lieberman, H.~Samet, and J.~Sankaranarayanan, ``Geotagging with local
  lexicons to build indexes for textually-specified spatial data,'' in
  \emph{Proc. IEEE Int. Conf. on Data Eng.}, 2010, pp. 201--212.

\bibitem{dill2003semtag}
S.~Dill, N.~Eiron, D.~Gibson, D.~Gruhl, R.~Guha, A.~Jhingran, T.~Kanungo,
  S.~Rajagopalan, A.~Tomkins, J.~A. Tomlin \emph{et~al.}, ``Semtag and seeker:
  Bootstrapping the semantic web via automated semantic annotation,'' in
  \emph{Proc. Conf. World Wide Web}, 2003, pp. 178--186.

\bibitem{mihalcea2007wikify}
R.~Mihalcea and A.~Csomai, ``Wikify!: linking documents to encyclopedic
  knowledge,'' in \emph{Proc. {ACM} Conf. on Information and Knowledge
  Management}, 2007, pp. 233--242.

\bibitem{milne2008learning}
D.~Milne and I.~H. Witten, ``Learning to link with wikipedia,'' in \emph{Proc.
  {ACM} Conf. on Information and Knowledge Management}, 2008, pp. 509--518.

\bibitem{witten2008effective}
I.~Witten and D.~Milne, ``An effective, low-cost measure of semantic
  relatedness obtained from wikipedia links,'' in \emph{Proc. of AAAI Workshop
  on Wikipedia and Artificial Intelligence: an Evolving Synergy}, 2008, pp.
  25--30.

\bibitem{kulkarni2009collective}
S.~Kulkarni, A.~Singh, G.~Ramakrishnan, and S.~Chakrabarti, ``Collective
  annotation of wikipedia entities in web text,'' in \emph{Proc. ACM Conf. on
  Knowledge Discovery and Data Mining}, 2009, pp. 457--466.

\bibitem{cucerzan2007large}
S.~Cucerzan, ``Large-scale named entity disambiguation based on wikipedia
  data,'' in \emph{Proc. Joint Conf. on Empirical Methods in Natural Language
  Processing and Computational Natural Language Learning}, 2007, pp. 708--716.

\bibitem{hoffart2011robust}
J.~Hoffart, M.~A. Yosef, I.~Bordino, H.~F{\"u}rstenau, M.~Pinkal, M.~Spaniol,
  B.~Taneva, S.~Thater, and G.~Weikum, ``Robust disambiguation of named
  entities in text,'' in \emph{Proc. Conf. on Empirical Methods in Natural
  Language Processing}.\hskip 1em plus 0.5em minus 0.4em\relax Association for
  Computational Linguistics, 2011, pp. 782--792.

\bibitem{DBLP:conf/sigir/HanSZ11}
X.~Han, L.~Sun, and J.~Zhao, ``Collective entity linking in web text: a
  graph-based method,'' in \emph{Proc. {ACM} {SIGIR} Conf. on Research and
  Development in Information Retrieval}, 2011, pp. 765--774.

\bibitem{tsochantaridis2005large}
I.~Tsochantaridis, T.~Joachims, T.~Hofmann, and Y.~Altun, ``Large margin
  methods for structured and interdependent output variables,'' \emph{J.
  Machine Learning Research}, vol.~6, no. Sep, pp. 1453--1484, 2005.

\bibitem{DBLP:conf/acl/ZhangC08}
Y.~Zhang and S.~Clark, ``Joint word segmentation and {POS} tagging using a
  single perceptron,'' in \emph{Proc. Meeting of the Association for
  Computational Linguistics}, 2008, pp. 888--896.

\bibitem{jeh2003scaling}
G.~Jeh and J.~Widom, ``Scaling personalized web search,'' in \emph{Proc. Conf.
  on World Wide Web}, 2003, pp. 271--279.

\bibitem{antoine2015portraying}
{\'E}.~Antoine, A.~Jatowt, S.~Wakamiya, Y.~Kawai, and T.~Akiyama, ``Portraying
  collective spatial attention in twitter,'' in \emph{Proc. {ACM} Conf. on
  Knowledge Discovery and Data Mining}, 2015, pp. 39--48.

\bibitem{finin2010annotating}
T.~Finin, W.~Murnane, A.~Karandikar, N.~Keller, J.~Martineau, and M.~Dredze,
  ``Annotating named entities in twitter data with crowdsourcing,'' in
  \emph{Proc. of the NAACL HLT 2010 Workshop on Creating Speech and Language
  Data with Amazon's Mechanical Turk}.\hskip 1em plus 0.5em minus 0.4em\relax
  Association for Computational Linguistics, 2010, pp. 80--88.

\bibitem{gelernter2011geo}
J.~Gelernter and N.~Mushegian, ``Geo-parsing messages from microtext,''
  \emph{Trans. GIS}, vol.~15, no.~6, pp. 753--773, 2011.

\bibitem{liu2014automatic}
F.~Liu, M.~Vasardani, and T.~Baldwin, ``Automatic identification of locative
  expressions from social media text: A comparative analysis,'' in \emph{Proc.
  Workshop on Location and the Web}, 2014, pp. 9--16.

\bibitem{derczynski2015analysis}
L.~Derczynski, D.~Maynard, G.~Rizzo, M.~van Erp, G.~Gorrell, R.~Troncy,
  J.~Petrak, and K.~Bontcheva, ``Analysis of named entity recognition and
  linking for tweets,'' \emph{Information Processing \& Management}, vol.~51,
  no.~2, pp. 32--49, 2015.

\bibitem{DBLP:conf/wsdm/DalviKP12}
N.~N. Dalvi, R.~Kumar, and B.~Pang, ``Object matching in tweets with spatial
  models,'' in \emph{Proc. Conf. on Web Search and Data Mining}, 2012, pp.
  43--52.

\bibitem{zhao2016annotating}
K.~Zhao, G.~Cong, and A.~Sun, ``Annotating points of interest with geo-tagged
  tweets,'' in \emph{Proc. ACM Conf. on Information and Knowledge Management},
  2016, pp. 417--426.

\vfill\eject

\bibitem{cheng2011exploring}
Z.~Cheng, J.~Caverlee, K.~Lee, and D.~Z. Sui, ``Exploring millions of
  footprints in location sharing services,'' in \emph{Proc. Int. Conf. on
  Weblogs and Social Media}, 2011.



\bibitem{ye2010location}
M.~Ye, P.~Yin, and W.-C. Lee, ``Location recommendation for location-based
  social networks,'' in \emph{Proc. SIGSPATIAL Int. Conf. on Advances in
  Geographic Information Systems}, 2010, pp. 458--461.

\bibitem{gao2012exploring}
H.~Gao, J.~Tang, and H.~Liu, ``Exploring social-historical ties on
  location-based social networks,'' in \emph{Proc. Int. Conf. on Weblogs and
  Social Media}, 2012.

\bibitem{gao2012gscorr}
------, ``{gSCorr}: modeling geo-social correlations for new check-ins on
  location-based social networks,'' in \emph{Proc. ACM Conf. on Information and
  knowledge management}, 2012, pp. 1582--1586.

\bibitem{yang2013sentiment}
D.~Yang, D.~Zhang, Z.~Yu, and Z.~Wang, ``A sentiment-enhanced personalized
  location recommendation system,'' in \emph{Proc. ACM Conf. on Hypertext and
  Social Media}, 2013, pp. 119--128.

\bibitem{gao2015content}
H.~Gao, J.~Tang, X.~Hu, and H.~Liu, ``Content-aware point of interest
  recommendation on location-based social networks,'' in \emph{Proc. {AAAI}
  Conf. on Artificial Intelligence}, 2015, pp. 1721--1727.

\bibitem{liu2013point}
B.~Liu and H.~Xiong, ``Point-of-interest recommendation in location based
  social networks with topic and location awareness,'' in \emph{Proc. {SIAM}
  Int. Conf. on Data Mining}, 2013, pp. 396--404.

\bibitem{liu2013learning}
B.~Liu, Y.~Fu, Z.~Yao, and H.~Xiong, ``Learning geographical preferences for
  point-of-interest recommendation,'' in \emph{Proc. ACM SIGKDD Conf. on
  Knowledge Discovery and Data Mining}, 2013, pp. 1043--1051.

\bibitem{cheng2013you}
C.~Cheng, H.~Yang, M.~R. Lyu, and I.~King, ``Where you like to go next:
  Successive point-of-interest recommendation,'' in \emph{Proc. Int. Joint
  Conf. on Artificial Intelligence}, 2013, pp. 2605--2611.

\bibitem{DBLP:conf/cikm/YaoZHB17}
D.~Yao, C.~Zhang, J.~Huang, and J.~Bi, ``{SERM:} {A} recurrent model for next
  location prediction in semantic trajectories,'' in \emph{Proc. {ACM} Conf. on
  Information and Knowledge Management}, 2017, pp. 2411--2414.

\bibitem{gao2014personalized}
H.~Gao, ``Personalized {POI} recommendation on location-based social
  networks,'' Ph.D. dissertation, Arizona State University, 2014.

\bibitem{ying2012urban}
J.~J.-C. Ying, E.~H.-C. Lu, W.-N. Kuo, and V.~S. Tseng, ``Urban
  point-of-interest recommendation by mining user check-in behaviors,'' in
  \emph{Proc. ACM SIGKDD Workshop on Urban Computing}, 2012, pp. 63--70.

\bibitem{cho2011friendship}
E.~Cho, S.~A. Myers, and J.~Leskovec, ``Friendship and mobility: user movement
  in location-based social networks,'' in \emph{Proc. ACM Conf. on Knowledge
  Discovery and Data Mining}, 2011, pp. 1082--1090.

\bibitem{DBLP:conf/kdd/HasanZU13}
S.~Hasan, X.~Zhan, and S.~V. Ukkusuri, ``Understanding urban human activity and
  mobility patterns using large-scale location-based data from online social
  media,'' in \emph{Proc. {ACM} {SIGKDD} Workshop on Urban Computing}, 2013,
  pp. 6:1--6:8.

\bibitem{bao2015recommendations}
J.~Bao, Y.~Zheng, D.~Wilkie, and M.~Mokbel, ``Recommendations in location-based
  social networks: a survey,'' \emph{GeoInformatica}, vol.~19, no.~3, pp.
  525--565, 2015.

\bibitem{zhao2016survey}
S.~Zhao, I.~King, and M.~R. Lyu, ``A survey of point-of-interest recommendation
  in location-based social networks,'' \emph{CoRR}, 2016.

\bibitem{Liu:2017:EEP}
Y.~Liu, T.~Pham, G.~Cong, and Q.~Yuan, ``An experimental evaluation of
  point-of-interest recommendation in location-based social networks,''
  \emph{{PVLDB}}, vol.~10, no.~10, pp. 1010--1021, 2017.

\end{thebibliography}
\end{document}